\documentclass{mn}
\usepackage[T1]{fontenc}
\usepackage{graphics}
\def\ltsima{$\; \buildrel < \over \sim \;$}
\def\simlt{\lower.5ex\hbox{\ltsima}}
\def\gtsima{$\; \buildrel > \over \sim \;$}
\def\simgt{\lower.5ex\hbox{\gtsima}}

\makeatletter

\providecommand{\LyX}{L\kern-.1667em\lower.25em\hbox{Y}\kern-.125emX\@}
\newcommand{\noun}[1]{\textsc{#1}}
\let\SF@@footnote\footnote
\def\footnote{\ifx\protect\@typeset@protect
    \expandafter\SF@@footnote
  \else
    \expandafter\SF@gobble@opt
  \fi
}
\expandafter\def\csname SF@gobble@opt \endcsname{\@ifnextchar[
  \SF@gobble@twobracket
  \@gobble
}
\edef\SF@gobble@opt{\noexpand\protect
  \expandafter\noexpand\csname SF@gobble@opt \endcsname}
\def\SF@gobble@twobracket[#1]#2{}

\usepackage{epsfig}

\usepackage[T1]{fontenc}



\begin{document}

\title{The circumnuclear X--ray reflectors in NGC~1068 and the Circinus Galaxy}

\author{Stefano Bianchi\protect\( ^{1}\protect \), Giorgio Matt\protect\( ^{1}\protect \),
Kazushi Iwasawa\protect\( ^{2}\protect \) \\
 ~\\
\protect\( ^{1}\protect \) Dipartimento di Fisica ``E.Amaldi'', Università
degli Studi ``Roma Tre'', Via della Vasca Navale 84, I-00146 Roma, Italy \\
 \protect\( ^{2}\protect \) Institute of Astronomy, University of Cambridge,
Madingley Road, Cambridge CB3 0HA, United Kingdom \\
 }

\maketitle
\begin{abstract}
The $ASCA$ and $BeppoSAX$ spectra of the Circinus Galaxy and NGC~1068 are
analysed and compared with photoionization models based on  {\sc cloudy}.
In the case of Circinus, a single, mildly ionized reflector can account for the
line spectrum, while in NGC~1068 at least three different reflectors (with
different ionization states) are needed. We suggest that
the reflector in Circinus, and the low ionized one in NGC~1068 are the inner
and visible part of the material responsible for the X--ray absorption. With
this assumption, we estimate for the inner radius of the absorber a value 
of 0.2 pc for Circinus and of a few parsecs for NGC~1068. 
\end{abstract}

\author{Stefano Bianchi, et al.}

\section{Introduction}

The heavy obscuration of the nuclear X--ray emission in Seyfert 2 galaxies
allows to observe and study components that in type 1 sources are diluted into
invisibility by the nuclear radiation. Among these components, particularly
interesting are those due to reflection of the nuclear X--rays from circumnuclear
matter, as they permit in principle to map the reflecting regions and probe
their physical and chemical properties through the study of the continuum and
especially of the emission line spectra (e.g. Netzer 1996; Matt, Brandt \& Fabian
1996; Krolik 1999 and references therein).

If the absorbing matter is Compton--thick (i.e. \( N_{H}\geq \sigma _{T}^{-1}=1.5\times10 ^{24} \)
cm\( ^{-2} \), see Matt et al. 2000 for the discussion of the general properties
of Compton--thick sources), the nuclear radiation is obscured up to at least
\( \sim  \)10 keV, and all astrophysically important X--ray lines (including
iron lines) emitted by the reflecting matter can be studied.

In this paper we discuss the $ASCA$ and $BeppoSAX$ X--ray spectra 
of the two brightest and best studied Compton--thick sources, i.e. the Circinus
Galaxy and NGC~1068. 
In the case of NGC~1068, we present for the first
time results from the AO4 100 ks \emph{ASCA} observation (results published so
far by Ueno et al. 1994, Iwasawa, Fabian \& Matt 1997, and Netzer \& Turner
1997, were obtained with the PV 40 ks observation). Our aim is to derive the
properties of the reflectors in these sources. To achieve this goal, we have
extensively used the photoionization code \textsc{cloudy}\footnote{
http://www.pa.uky.edu/\~\,gary/cloudy/} (version 90.05; Ferland 1997), to construct theoretical models to be compared 
with the observations.

The paper is organized as follows: Sec. 2 summarizes previous X--ray observations
of the two sources; Sec. 3 describes the data reduction, while Sec. 4 reports
the spectral analysis and Sec. 5 compare the results with the theoretical models.
In each section we will discuss first the Circinus Galaxy and then NGC~1068.
Finally, the results will be summarized in Sec. 6.

\section{Previous X--ray observations}

\subsection{The Circinus Galaxy}

The Circinus Galaxy, one of the closest AGN and X--ray brightest Seyfert 2 galaxies,
was observed for the first time in X--rays during the \emph{ROSAT} All Sky Survey
\cite{bsb}. Later on, it was observed by \emph{ASCA}; the spectrum is dominated
by a pure Compton reflection component, with a very prominent iron K$\alpha$
line and several other lines from lighter elements (Matt et al. 1996). The 
lines from the lighter elements have energies indicating significant ionization
of the emitting material (Matt et al. 1996; Netzer et al. 1998; Sako et al. 
2000). 
The \emph{BeppoSAX}
observation, besides confirming the \emph{ASCA} findings below 10 keV, detected
the nuclear emission, seen through very thick (4$\times10^{24}$ cm$^{-2}$)
absorbing matter (Matt et al. 1999; Guainazzi et al. 1999).

\subsection{NGC 1068}

The history of the X--ray observations of NGC~1068 is longer and richer than that of the Circinus
Galaxy. The source was observed by \textit{Einstein} (Monier \& Halpern 1987)
and \emph{EXOSAT} (Elvis \& Lawrence 1988); the lack of any measured intrinsic absorption
was interpreted by those authors as evidence for reflected emission of an obscured
nucleus. This scenario was confirmed by the \textit{Ginga} detection of a strong
iron line (Koyama et al. 1989). At softer energies, the ROSAT HRI discovery
of a very extended (\( \sim  \)13 kpc) emission indicated that a large part
of the radiation in this band is related to the host galaxy rather than the nucleus.

\emph{ASCA} (Ueno et al. 1994; Iwasawa, Fabian \& Matt 1997; Netzer \& Turner
1997) confirmed both the thermal--like emission in soft X--rays and the reflection
component at higher energies. The very rich line spectrum and the complex iron
line (including a neutral or low ionized line as well as He-- and H--like lines)
indicate that more than one reflector is present. Smith, Done \& Pounds (1993)
proposed to identify the cold reflector with the inner surface of the putative
molecular torus, an idea further pursued by Matt, Brandt \& Fabian (1996) and
Iwasawa, Fabian \& Matt (1997). \emph{BeppoSAX} confirmed the complexity of
the reflection in NGC~1068 by separating at least two reflecting continua, one
cold and one ionized (Matt et al. 1997; Guainazzi et al. 1999). The lack of
any transmitted component implies that the absorbing matter should have a column
density exceeding 10\( ^{25} \) cm\( ^{-2} \).

Finallly,  evidence for flux and spectral variability between the two \emph{BeppoSAX} observations
is discussed by Guainazzi et al. (2000).

\section{Data reduction}

\subsection{The Circinus Galaxy}

In the spectral analysis discussed below we will mainly use \emph{ASCA} data
(observation performed on 1995, August 14), but we will also take advantage
of the \emph{BeppoSAX} results (observation performed on 1998, March 24).

Details on the \emph{BeppoSAX} observation and related data reduction can be
found in Matt et al. (1999) and Guainazzi et al. (1999).
Regarding $ASCA$, 
we have verified the previous results (Matt et al. 1996) using the Revision2 data
processing and the latest calibration. It should be noted that the $SIS0$
spectrum appears to be affected by a gain problem: e.g., the iron K line
at 6.4 keV is shifted to a higher energy (see below for details) and
a similar energy shift is seen for the instrumental Ni K$\alpha$ line at
7.5 keV. 
We have checked the instrumental Ni K$\alpha $ line at 7.5 keV for the two $SIS$
data taken from a source-free region on the detector field of view and 
found that
the $SIS0$ data show an energy shift similar to that seen in the source spectrum
while the $SIS1$ data show the line at the right energy. We therefore believe 
that
the line energies derived from the SIS1 are more reliable. Also, a recent 
$Chandra$
HETG observation of the Circinus galaxy found a narrow line at 6.4 keV 
(R. Sambruna, private  communication).

\subsection{NGC~1068}

In the spectral analysis discussed below we have used both \emph{ASCA} (observation
performed on 1996, August 15) and \emph{BeppoSAX} (observations performed on
1996, December 30 and 1998, January 11) data.

Details on the \emph{BeppoSAX} observations and related data reduction can be
found in Matt et al. (1997) and Guainazzi et al. (1999, 2000).

The $ASCA$ data presented here are from a long observation carried out in
the AO-4. The data reduction was performed in the standard procedure using
FTOOLS version 4.2 and the latest calibration available. The exposure time
for both $SIS$ detectors is 118 ks, respectively.

\section{Spectral analysis}

\subsection{The Circinus Galaxy}

\begin{figure}
{\par\centering \resizebox*{1\columnwidth}{!}{\rotatebox{-90}{\includegraphics{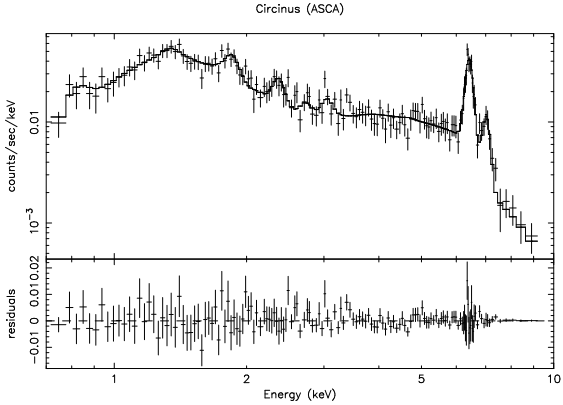}}} \par}

\caption{\label{circascafit}\emph{ASCA SIS1} data and best model for Circinus.}
\end{figure}

\begin{figure}
{\par\centering \resizebox*{1\columnwidth}{!}{\rotatebox{-90}{\includegraphics{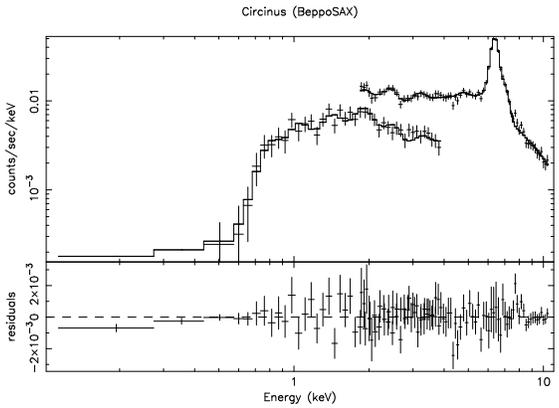}}} \par}

\caption{\label{circsaxfit}\emph{BeppoSAX LECS+MECS} data and best fit model for
Circinus.}
\end{figure}

We have re--analysed the \emph{ASCA} \emph{SIS1} spectra of the Circinus Galaxy
to search for emission lines, taking advantage of the results on the continuum
obtained with \emph{BeppoSAX} \cite{m99}. We have fitted the spectrum with
the following model:

\begin{equation}
F=\left[ R_{C}+PL_{S}+emissionlines\right] \exp \{-\sigma _{ph}N_{H}\}
\end{equation}

\noindent where \( R_{C} \) is the cold reflection component 
(model \textsc{pexrav};
the power law photon index of the nuclear illuminating continuum has been fixed
to the value of 1.56 obtained with \emph{BeppoSAX}, Matt et al. 1999), \( PL_{S} \)
is a power law component describing the soft X--ray emission, and \( N_{H} \)
is the cold absorption along the line of sight, which includes absorption from
our own Galaxy plus possible contributions from the Circinus Galaxy itself.
The primary, heavily absorbed continuum discovered by \emph{BeppoSAX} has not
been included in the model because it does not affect the \emph{ASCA} energy
band. We have added as many emission lines as required by the data, inspecting
by eye the residuals after any single addition. All lines have been assumed
to be narrow for simplicity, even if, as discussed in Sec.~\ref{res_circ},
some of them are very likely a blend of different lines with comparable fluxes.
With all lines included, the reduced \( \chi ^{2} \) is 0.8 for 141 d.o.f..
The best fit parameters for the continuum are listed in Table \ref{Circresume},
while Fig. \ref{circascafit} and \ref{circsaxfit} show the best fit models
for \emph{ASCA} and \emph{BeppoSAX}. 

We will focus our interest on the line spectrum in Sec. \ref{res_circ}:
the best fit line parameters are listed in Table \ref{circobsline}, along with
the most likely identifications. \emph{SIS1} data are systematically different
from those of \emph{SIS}0: the line energies (even if always at least marginally
consistent with the other instrument) are all shifted to higher values, while
the equivalent widths (except for the first one) are all much lower than those
of \emph{SIS0}. The reason for this discrepancy is not clear. 
The EWs measured by \emph{SIS0} may be more reliable, since this instrument
is considered to be best calibrated at lower energies,
while the correct line energies are probably those measured by \emph{SIS1}
(see discussion in Sec.~3.1). The value of the reduced \( \chi ^{2} \)
for the fit of the continuum is similar for the two instruments.

Upper limits on the equivalent widths of the ionized iron lines are 30 eV 
for the 6.7 keV Fe\( \,  \)\textsc{xxv} line, and 340 eV for the 6.96  
Fe\( \,  \)\textsc{xxvi} line.

The photon index of the soft power law is not very well constrained, i.e. {\small \( \Gamma _{s}=2.66_{-0.43}^{+0.62} \)},
while the column density of the absorbing matter is 6.6(\( \pm  \)2.1)\( \times 10^{21} \)
cm\( ^{-2} \). (The fit to the \emph{BeppoSAX} data, after fixing the same
value for the column density, gives a flatter power law with \( \Gamma _{s}=1.75_{-0.92}^{+0.28} \)).
A thermal plasma model for the soft excess does not provide a better fit to
the data. The temperature is 1.84\( ^{+1.20}_{-0.41} \) keV, while the abundance
is constrained to be less than 0.7 times the solar value. As the thermal model
cannot account for the observed lines, we do not discuss it any further.
The 0.5--2 keV luminosity of the soft component is around 7$\times10^{39}$ 
erg cm$^{-2}$ s$^{-1}$. The IR/soft--X--ray ratio is similar to that of the
sources in the sample of powerful infrared galaxies studied by Iwasawa (1999).

\begin{table*}

\begin{minipage}{175mm}

\caption{Best fit parameters for the continuum of Circinus. The values with the symbol \( ^{*} \) were kept fixed during the fit.}

\label{Circresume}

{\centering \begin{tabular}{|c||cc|cc|}
\hline 
&
\multicolumn{2}{|c|}{\textbf{\emph{ASCA}} \textbf{}}&
\multicolumn{2}{|c|}{\textbf{\emph{BeppoSAX}} }\\
\cline{2-3} \cline{4-5} 
&
\textsf{\emph{\noun{\footnotesize powerlaw}}}&
\textsf{\emph{\noun{\footnotesize mekal}}}&
\textsf{\emph{\noun{\footnotesize powerlaw}}}&
\textsf{\emph{\noun{\footnotesize mekal}}}\\
\hline 
\hline 
\emph{reduced \( \chi ^{2} \)} &
0.8 (141 d.o.f.)&
0.8 (140 d.o.f.)&
1 (92 d.o.f.)&
1 (91 d.o.f.)\\
\hline 
\emph{\( \Gamma  \)}&
\( 1.56^{*} \)&
\( 1.56^{*} \)&
\( 1.56^{*} \)&
\( 1.56^{*} \)\\
\hline 
\emph{cut-off (keV)}&
\( 56^{*} \)&
\( 56^{*} \)&
\( 56^{*} \)&
\( 56^{*} \)\\
\hline 
\emph{\( N_{H} \) (\( cm^{-2} \))} &
{\small \( \left( 6.6_{-2.1}^{+2.1}\right) \times 10^{21} \)}&
{\small \( \left( 5.7_{-1.7}^{+2.3}\right) \times 10^{21} \)}&
\( 6.6\times 10^{21}\, ^{^{*}} \) &
\( 6.6\times 10^{21}\, ^{^{*}} \)\\
\hline 
\emph{\( \Gamma _{s} \)}&
{\small \( 2.66_{-0.43}^{+0.62} \)}&
-&
{\small \( 1.75_{-0.92}^{+0.28} \)}&
-\\
\hline 
\emph{\( kT \) (keV)}&
-&
{\small \( 1.84_{-0.41}^{+1.2} \)}&
-&
{\small \( 3.49_{-0.75}^{+1.1} \)}\\
\hline 
\emph{\( A_{Z} \)}&
-&
{\small \( \left( 1.3_{-1.2}^{+5.5}\right) \times 10^{-1} \)}&
-&
{\small \( 2.3_{-1.6}^{+6.52} \)}\\
\hline 
\end{tabular}\small \par}

\end{minipage}

\end{table*}

\subsection{NGC~1068}

As already noted by Iwasawa, Fabian \& Matt (1997) and Guainazzi et al. (1999),
the X--ray spectrum of NGC~1068 is much more complex than that of
the Circinus Galaxy. The continuum is composed by at least three components:
a thermal--like one, probably associated with the extended starburst observed
by \emph{ROSAT} \cite{w92} and dominating the soft X--ray band; and at least
two reflectors, one cold and one ionized (Iwasawa, Fabian \& Matt 1997, Matt
et al. 1997). Superimposed on these components are many emission lines, most
of them associated with the ionized reflector(s), even if the neutral K\( \alpha  \)
line is clearly visible.

We adopted for the continuum the same model of Guainazzi et al. (1999): a
single temperature thermal plasma model (\textsc{mekal} in \textsc{Xspec}),
a cold reflection component (\textsc{pexrav}) and an ionized reflector component,
assumed to be a simple power law with the same spectral index of the power law
illuminating the cold matter, which in turn has been fixed to 2.13, as measured
by \emph{BeppoSAX} \cite{g99}. The column density of the absorbing matter has
been fixed to the Galactic value of 3.5\( \times  \)10\( ^{20} \) cm\( ^{-2} \)
(Dickey \& Lockman 1990). Then, as in Circinus, we added emission lines until
required by the data. The best fit parameters for the continuum are given in
Table \ref{NGCresume}, giving a reduced \( \chi ^{2} \) of \( \simeq 1.1 \)
for 582 d.o.f.. They are consistent with those obtained by Guainazzi et al.
(1999) from the \emph{BeppoSAX} data. 

The free parameters in the fit, apart from the energies and the normalizations
of the lines, are the temperature and the abundance
of the \textsc{mekal} model: \( T=0.7\pm 0.1\, keV \) and \( A_{Z}=\left( 1.3_{-0.2}^{+0.1}\right) \times 10^{-1} \).

\begin{figure}
{\par\centering \resizebox*{1\columnwidth}{!}{\rotatebox{-90}{\includegraphics{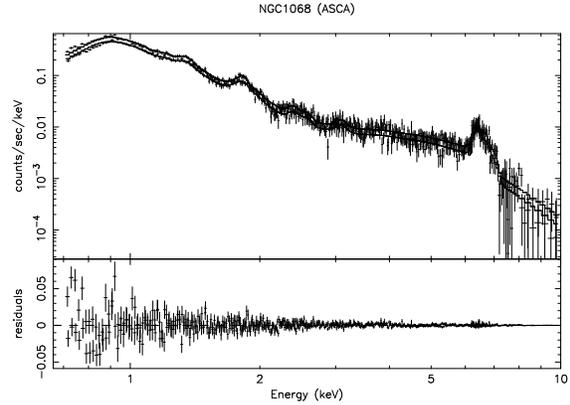}}} \par}

\caption{\label{1068ascafit}\emph{ASCA SIS0+SIS1} data and best fit model for NGC
1068.}
\end{figure}

\begin{figure}
{\par\centering \resizebox*{1\columnwidth}{!}{\rotatebox{-90}{\includegraphics{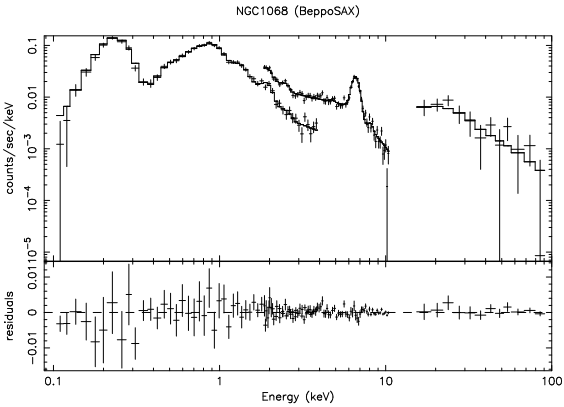}}} \par}

\caption{\label{1068saxfit}\emph{BeppoSAX LECS+MECS+PDS} data and best fit 
model for NGC 1068.}
\end{figure}

A fit to the \emph{BeppoSAX} data (Fig. \ref{1068saxfit}) with the same model
gives a similar value for the reduced \( \chi ^{2} \), but the \textsc{mekal}
parameters are not consistent with the \emph{ASCA} ones: \( T=0.40^{+0.01}_{-0.01} \)
keV and \( A_{Z}=\left( 1.6_{-0.2}^{+0.3}\right) \times 10^{-2} \). Two more
lines are needed for the \emph{BeppoSAX} data with respect to those found with
\emph{ASCA}, \emph{}at energies of 0.55 and \( 8.2\, keV \). This is the
result of the larger band coverage of the instruments aboard \emph{BeppoSAX}.
The lines at 0.55 e \( 8.2\, keV \) have been both observed and discussed by
Guainazzi et al. (1999). The identification of the second one with the \( K\beta  \)
transition of Fe\( \,  \)\textsc{xxvi} at \( 8.2\, keV \) \cite{g99} allows 
to fix its energy and calculate the ratio against the \( K\alpha  \) of 
Fe\( \,  \)\textsc{xxvi}, which is \( 0.17_{-0.07}^{+0.10} \).

The line at \( 0.55_{-0.02}^{+0.03}\, keV \) confirms one of the most 
interesting results of Guainazzi et al. (1999), i.e. 
the detection for the first time  of the O\( \,  \)\textsc{vii}
line. It is worth remarking that this
line is in excess of what expected from the thermal--like component, and 
therefore it is likely to be produced in one of the reflectors. 
This line was not observed 
in previous observations because of the
low energy resolution (\emph{ROSAT}) or the narrow band coverage (\emph{ASCA}).
The upper limit set by the BBXRT observation 
suggested that the abundance of the oxygen could be less than the cosmic
one (Marshall et al. 1993; Netzer \& Turner 1997). Instead, the clear presence
of this line in the \emph{BeppoSAX} data overcomes this difficulty and, as we
will see in Sec. \ref{1068_res}, puts strong constrains to the number and the
characteristics of the reflectors in NGC 1068. 

\begin{table}

\caption{\label{NGCresume}Best fit parameters for the continuum of NGC 1068. The values
with the symbol \protect\( ^{*}\protect \) were kept fixed during the fit.}
{\centering \begin{tabular}{|c||c|c|}
\hline 
&
\textbf{\emph{ASCA}}&
\textbf{\emph{BeppoSAX}}\\
\hline 
\hline 
\emph{reduced \( \chi ^{2} \)}&
\( \simeq  \)1.1 (582 d.o.f.)&
\( \simeq  \)1.3 (111 d.o.f.)\\
\hline 
\emph{\( \Gamma  \)}&
\( 2.13^{*} \)&
\( 2.13^{*} \)\\
\hline 
\( \Gamma _{s} \)&
\( 2.13^{*} \)&
\( 2.13^{*} \)\\
\hline 
\emph{\( N_{H} \) (\( cm^{-2} \))} &
\( 3.5\times 10^{20}\, ^{^{*}} \) &
\( 3.5\times 10^{20}\, ^{^{*}} \)\\
\hline 
\emph{\( kT \) (keV)}&
\( 0.7\pm 0.1 \)&
{\small \( 0.40^{+0.01}_{-0.01} \)}\\
\hline 
\emph{\( A_{Z} \)}&
{\small \( \left( 1.3_{-0.2}^{+0.1}\right) \times 10^{-1} \)}&
{\small \( \left( 1.6_{-0.2}^{+0.3}\right) \times 10^{-2} \)}\\
\hline 
redshift&
\( 3.793\times 10^{-3}\, ^{^{*}} \)&
\( 3.793\times 10^{-3}\, ^{^{*}} \)\\
\hline 
\end{tabular}\par}\end{table}

All emission lines are listed in Tables \ref{NGCascaFelines} and \ref{NGCascalines}:
we will discuss them in detail in Sec. \ref{1068_res}.

\section{Comparison with models}

\subsection{\label{res_circ}The Circinus Galaxy}

\subsubsection{The model}

The unambiguous presence of a Compton reflection component in the \emph{ASCA}
(Matt et al. 1996) and \emph{BeppoSAX} \cite{m99} spectra indicate the presence
of thick circumnuclear matter. From the ratio between
the reflected and direct components,
it is likely that the covering factor of this matter is fairly large (Matt et
al. 1999, Guainazzi et al. 1999, Matt et al. 2000). It is then natural to assume
that the reflecting and absorbing media are one and the same. We have therefore
calculated with \textsc{cloudy} the reflected spectrum from a X--ray illuminated
material with a column density of 4.3\( \times  \)10\( ^{24} \) cm\( ^{-2} \)
\cite{m99}, adopting the dimensionless ionization parameter \( U \), defined
as:

\begin{equation}
U={Q\over 4\pi r^{2}cn}
\end{equation}

\noindent where \( Q \) is the total number of ionizing photons per unit time
reaching the surface of a cloud at a distance \( r \) from the source, while
\( n \) is the hydrogen number density. With the nuclear X--ray spectrum and
luminosity of Circinus \cite{m99}, we obtain (cgs units):

\begin{equation}
U\simeq {1.04\times10 ^{40}\over nr^{2}}
\end{equation}

\noindent From the measurements of the dynamical mass, Risaliti et al. (1999)
have set an upper limit of \( \sim  \)20 pc on the external radius of the cold
matter, \( R \), in agreement with recent estimates from IR observations, which
give a value of \( R\simeq 16\, pc \) \cite{r00}. Thus, assuming \( r\ll R \),
the density is \( \simeq 8.7\times10 ^{4} \) cm\( ^{-3} \), and the inner
surface of the cloud is \( r\simeq  \)0.11\( U^{-{1 \over 2}} \) pc. As described below,
we have found a solution with \( U \)\( \simeq  \)0.25, which implies a value
for \( r \) of $\sim$0.2 pc, roughly consistent with the water maser 
measurements
of Greenhill et al. (1997) and with the value of 0.5 pc estimated
from the IR observations whose \( R \) we adopted in our calculations. Then
the assumption \( r\ll R \) is, a posteriori, justified.
(It must be noted that the above determination of  \( r \) 
is based on the simplyfing assumption of constant density along the torus. 
A negative density gradient would, however, be probably more realistic. 
If this is the case, this value of \( r \) must be regarded 
as an upper limit.)

\begin{figure*} 

\begin{minipage}{175mm}

\vspace{0.3cm}
{\par\centering \resizebox*{0.45\columnwidth}{!}{\includegraphics{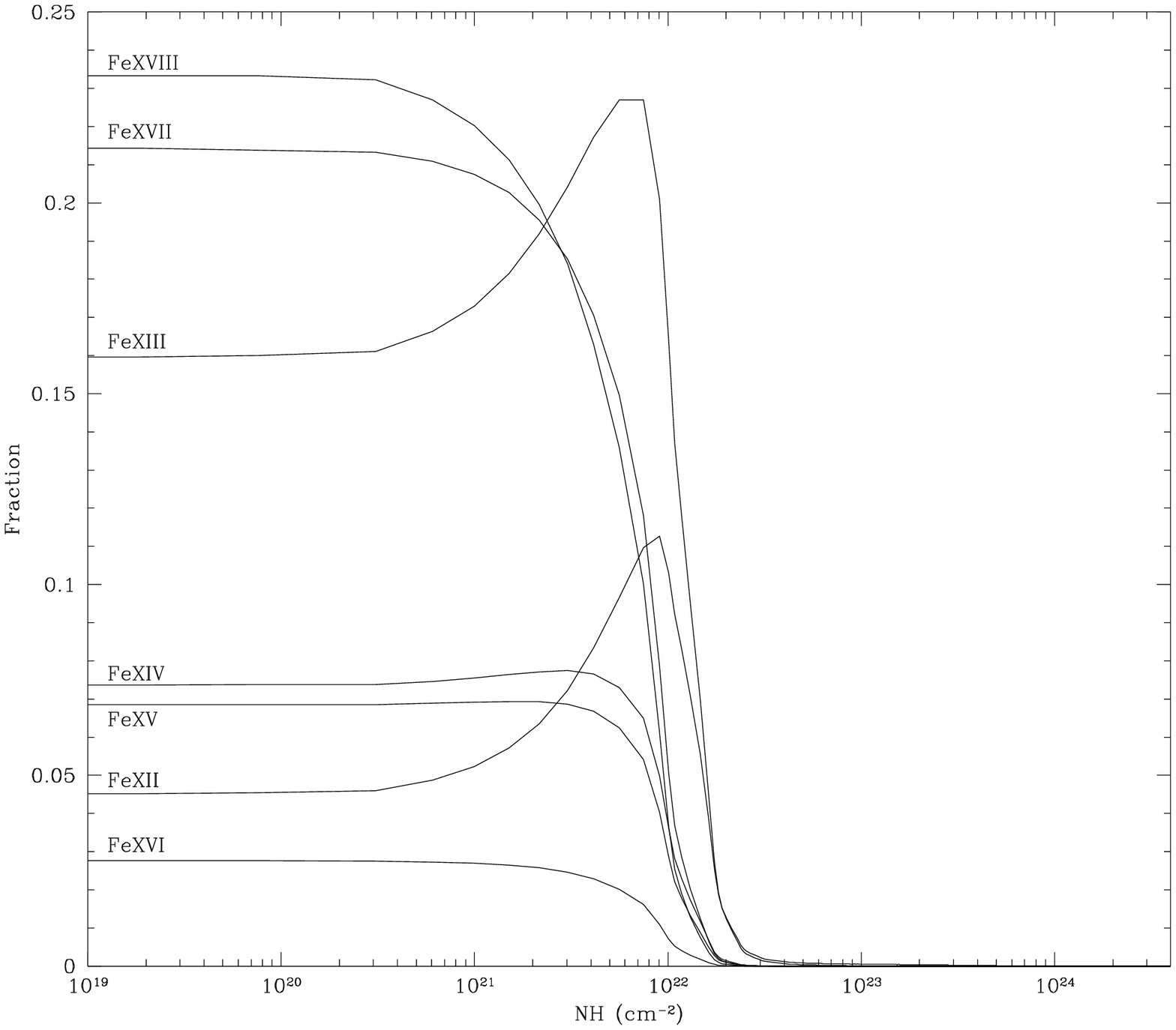}} 
\resizebox*{0.45\columnwidth}{!}{\includegraphics{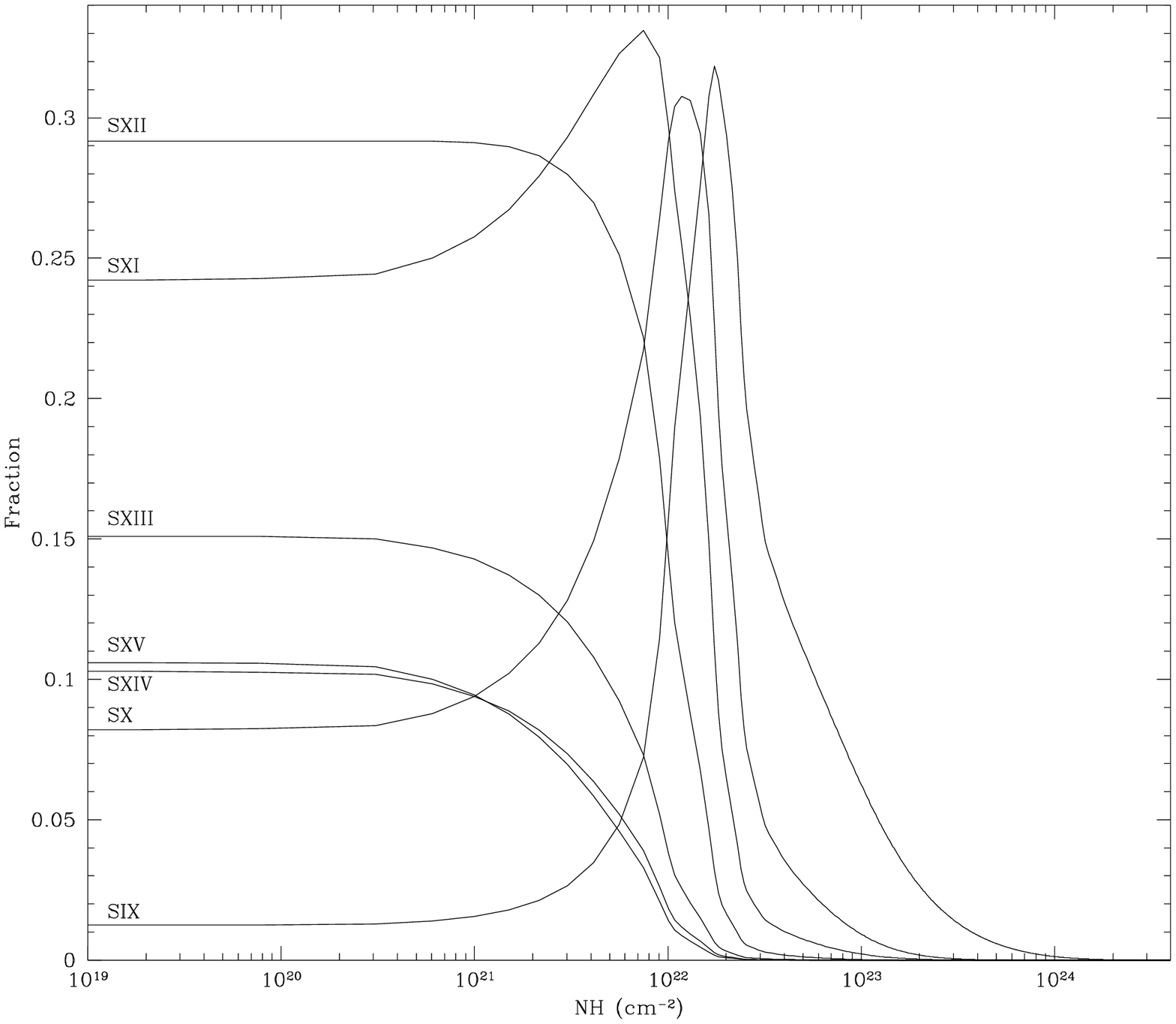}} \par}
\vspace{0.3cm}

\vspace{0.3cm}
{\par\centering \resizebox*{0.45\columnwidth}{!}{\includegraphics{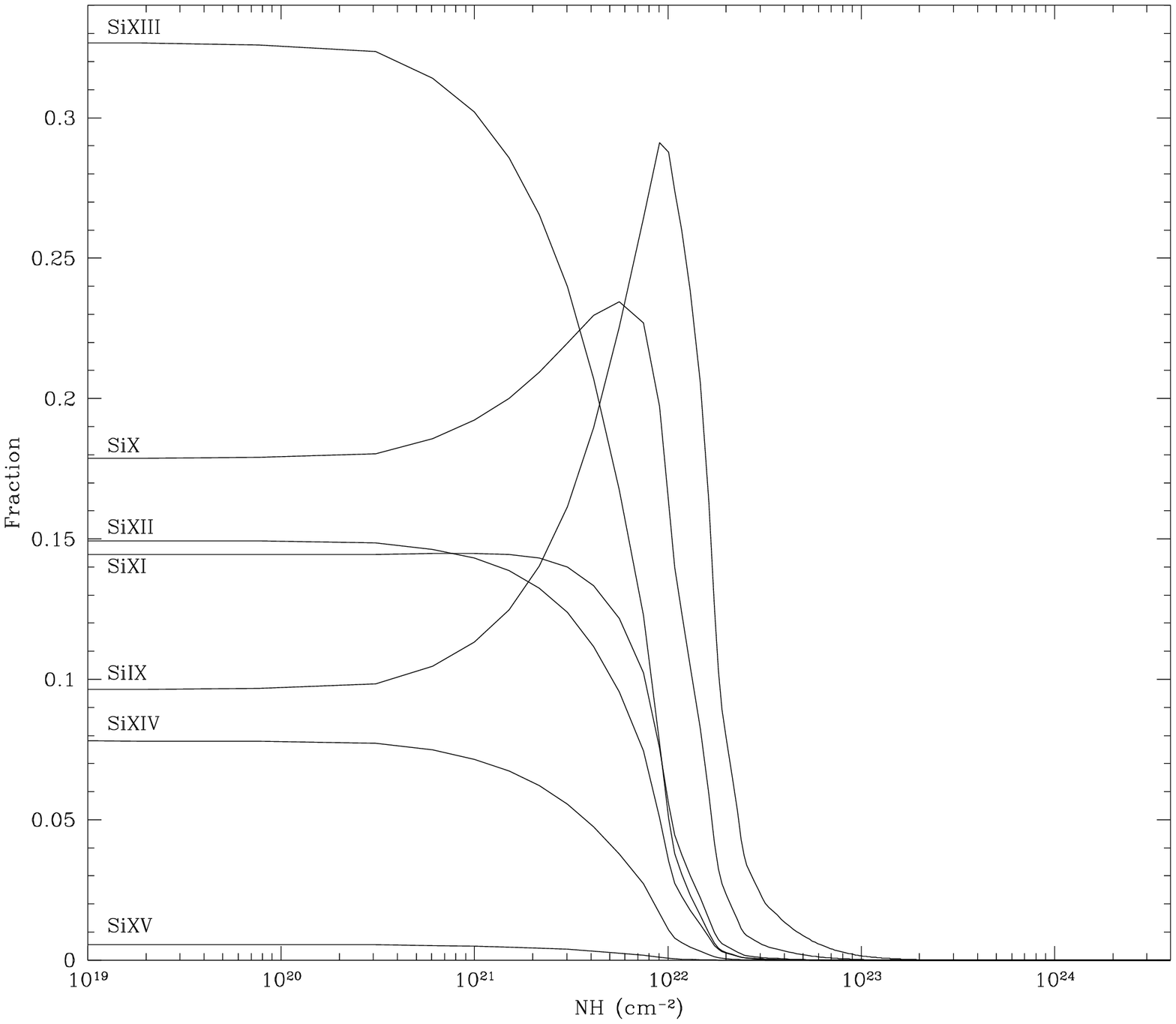}} 
\resizebox*{0.45\columnwidth}{!}{\includegraphics{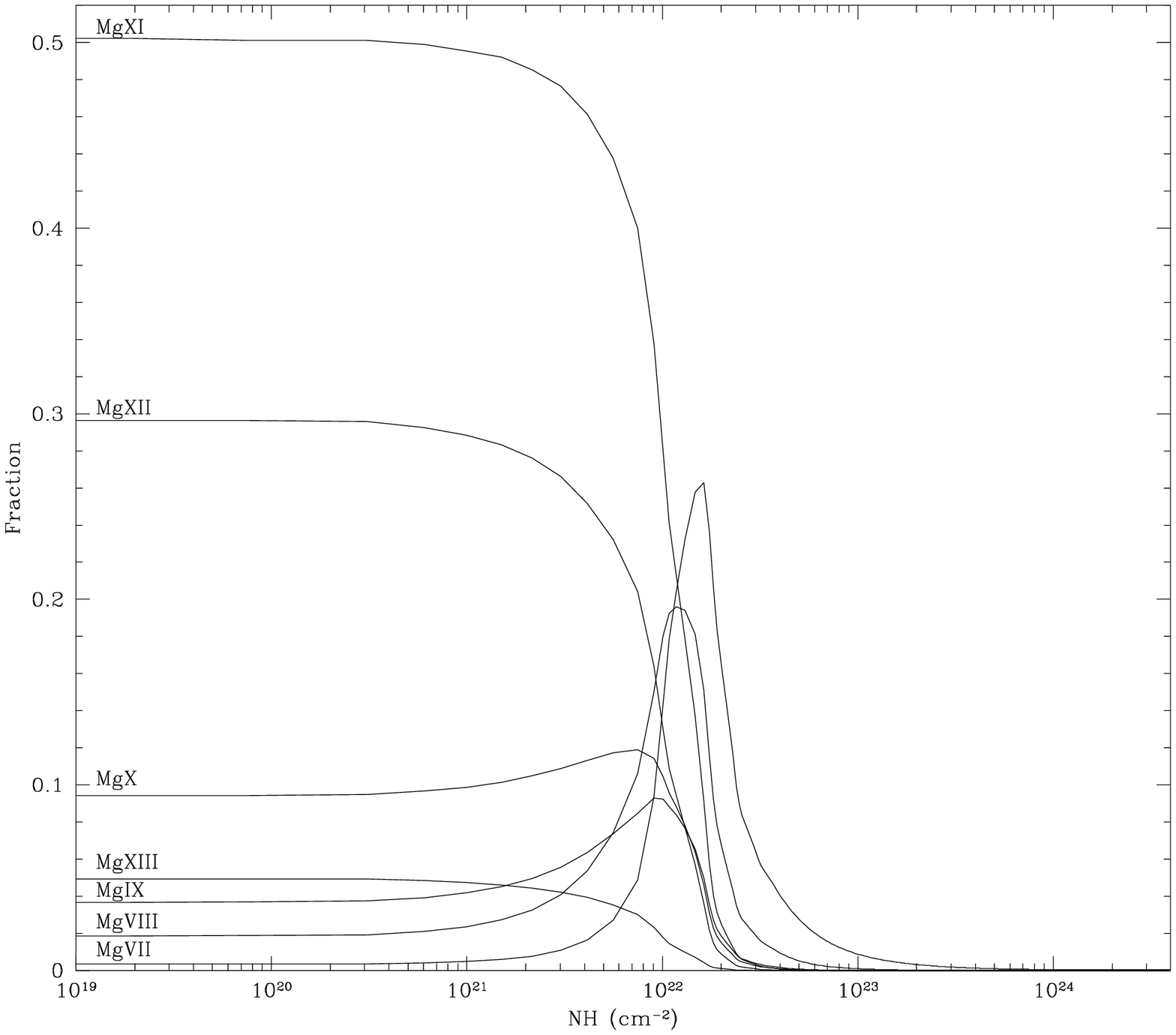}} \par}
\vspace{0.3cm}

\caption{Ionization structures in Circinus for Fe, S, Si and Mg. }

\label{circioniz}

\end{minipage}

\end{figure*}

We searched for the simplest possible solution, i.e. a single
reflector. The iron line undoubtedly comes from low ionized atoms, while
lighter elements are fairly ionized. We therefore searched for the highest value
of \( U \) consistent with iron less ionized than \textsc{xvii}, to ensure
that the iron line is still at \( \sim  \)6.4 keV, and that the resonant trapping
(e.g. Matt, Fabian \& Ross 1993) is not effective. We found such a solution
with \( U \)=0.25. The ionization structures for iron, sulphur, silicon and
magnesium are shown in Fig. \ref{circioniz}.

As a consistency check, we refitted the data with an ionized reflector 
(i.e. \textsc{pexriv} in \textsc{Xspec})
instead of the cold one. 
The ionization parameter used by \textsf{\footnotesize }\textsc{pexriv} has
a different definition:

\begin{equation}
\label{pexrivxi}
\xi \equiv \frac{L}{nr^{2}}\, erg\, cm\, s^{-1}
\end{equation}
 where \( L \) is the \( 5\, eV-20\, keV \) luminosity, assuming the continuum
as a simple power law (Magdziarz \& Zdziarski, 1995). The conversion factor
between the two definitions is \( \frac{\xi }{U}=249\, erg\, cm\, s^{-1} \).
Thus, the ionization parameter used in our model corresponds to \( \xi =249\cdot 0.25\simeq 62\, erg\, cm\, s^{-1} \). 

We first checked if reflection from ionized material 
could explain the soft excess, and then we fitted \emph{ASCA} and \emph{BeppoSAX}
data without the additional soft power law. In both cases, the 
fit is unacceptable, giving a reduced \( \chi ^{2} \) greater than 2. 
We then fitted the data including the power law component, and fixing 
\( \xi =62 \). The
ionization parameter has been fixed because it was not well constrained:
the value obtained with \textsc{cloudy} is more reliable, as \textsc{pexriv}
calculates the ionization equilibria in a simplified way (Done et al., 1992).
In these conditions, the \emph{ASCA} fit has a reduced \( \chi ^{2} \) almost
identical to the one obtained with \textsc{pexrav}. The soft power law photon
index is \( \Gamma _{s}=2.18_{-0.58}^{+0.52} \), while the absorbing column
density \( N_{H}=\left( 3.6_{-1.5}^{+1.5}\right) \times 10^{21}\, cm^{-2} \).
The flux in the power law is now a factor 2 lower than for the cold reflector.
The results for \emph{BeppoSAX} data are roughly the same, but again
with a photon index (\( \Gamma _{s}=0.71_{-0.07}^{+0.08} \)) which is not consistent
with the \emph{ASCA} one. 

Finally, we have estimated the solid angle of the reflecting matter which is
observable by comparing the observed 
\emph{}\( F_{o}=3.1(\pm 0.3)\times 10^{-4}\, ph\, cm^{-2}s^{-1} \) \( \simeq 3.2\times 10^{-12}erg\, cm^{-2}s^{-1} \)),
and calculated 
iron line fluxes. We find a value of 0.2$\times 4\pi$.


\subsubsection{The line spectrum}

\begin{table}

\caption{\label{circobsline}X-ray emission lines in Circinus from \emph{ASCA} data
(\emph{SIS1} and \emph{SIS0}). The EWs are calculated against the reflection
continuum (\textsc{pexriv}). All line profiles have been assumed to be 
\protect\( \delta \protect \)--functions.}
{\centering \begin{tabular}{|cc||cc|c|}
\hline 
\multicolumn{2}{|c||}{\textbf{\emph{SIS1}}\textbf{}}&
\multicolumn{2}{|c|}{\textbf{\emph{SIS0}} }&
\\
\hline 
\hline 
\multicolumn{1}{|c}{\textbf{E(keV)}}&
\textbf{EW(eV)}&
\textbf{E(keV)}&
\textbf{EW(eV)}&
\textbf{Id.}\\
\hline 
{\small \( 0.82_{-0.05}^{+0.05} \)}&
{\small \( 910_{-740}^{+910} \)}&
{\small \( 0.92_{-0.04}^{+0.04} \)}&
{\small \( 2500_{-1500}^{+2500} \)}&
Ne\( \,  \)\textsc{ix} {\small }\\
\hline 
{\small \( 1.34_{-0.05}^{+0.05} \)}&
{\small \( 645_{-180}^{+180} \)}&
{\small \( 1.38_{-0.03}^{+0.04} \)}&
{\small \( 530_{-310}^{+310} \)}&
Mg\( \,  \)\textsc{xi}{\small }\\
\hline 
{\small \( 1.85_{-0.02}^{+0.03} \)}&
{\small \( 910_{-240}^{+250} \)}&
{\small \( 1.87_{-0.02}^{+0.02} \)}&
{\small \( 790_{-190}^{+200} \)}&
Si\( \,  \)\textsc{xiii}{\small }\\
\hline 
{\small \( 2.36_{-0.03}^{+0.03} \)}&
{\small \( 580_{-370}^{+390} \)}&
{\small \( 2.40_{-0.04}^{+0.06} \)}&
{\small \( 305^{+130}_{-130} \)}&
Si\( \,  \)\textsc{xiii}{\small }\\
\hline 
{\small \( 2.72_{-0.32}^{+0.11} \)}&
{\small \( 130_{-120}^{+130} \)}&
{\small \( 3.00_{-0.30}^{+0.10} \)}&
{\small \( 60^{+80}_{-60} \)}&
S\( \,  \)\textsc{xv/xvi}{\small }\\
\hline 
{\small \( 3.05_{-0.07}^{+0.07} \)}&
{\small \( 170_{-110}^{+110} \)}&
{\small \( 3.22_{-0.07}^{+0.08} \)}&
{\small \( 150_{-100}^{+90} \)}&
Ar\( \,  \)\textsc{xvii}{\small }\\
\hline 
{\small \( 6.41_{-0.01}^{+0.02} \)}&
{\small \( 2130_{-220}^{+210} \)}&
{\small \( 6.49_{-0.01}^{+0.02} \)}&
{\small \( 1670_{-150}^{+150} \)}&
{\small Fe \( K\alpha  \)}\\
\hline 
{\small \( 7.00_{-0.04}^{+0.04} \)}&
{\small \( 640_{-170}^{+170} \)}&
{\small \( 6.98_{-0.27}^{+0.10} \)}&
{\small \( 260_{-100}^{+100} \)}&
Fe \( K\beta  \)\\
\hline 
\end{tabular}\par}\end{table}

The spectrum is rich in emission lines, and it is likely that in most cases
the observed lines are actually a blend of two or more lines with comparable
intensities. In Tab. \ref{circmodline} we report the observed equivalent width
of the lines (measured with respect to the reflected continuum only) compared
with the total EW of the most important lines likely contributing to any single
blend, as calculated by \textsc{cloudy}. The agreement between observed and
calculated one is good 
for the lines with energies above \( 1.85\, keV \) provided that the 
abundance is significantly higher than the solar one, 
i.e. \( A_{Z}\simgt 3 \) for the \emph{SIS0}
data (\( A_{Z}\simgt 4 \) for the less reliable \emph{SIS1} data)
(However, it should be noted that Oliva et al. (1999) found, instead, 
evidence for lower than solar metal abundances). 

\begin{table}

\caption{\label{circmodline}X-ray emission lines produced in our model for Circinus.
The energy in the first column is the mean value (weighted on the flux) of all
the lines that contribute to that blend. The last column shows the ratio between
the observed EW (\emph{SIS0} data) and that calculated by \textsc{cloudy}. All
the lines are \protect\( K\protect \) transitions. The 
\protect\( K\alpha \protect \)
lines which dominate each blend are in boldface.}
{\centering \begin{tabular}{|c|c|cc|c|}
\hline 
\textbf{\( \overline{E} \)(keV)}&
\textbf{EW(eV)}&
\textbf{Id.}&
\textbf{E(keV)}&
{\small \( \frac{EW_{0}}{EW_{C}} \)}\\
\hline 
\hline 
&
&
\textsc{Si\( \,  \)xiii}\textbf{}&
\textbf{1.85}&
\\
1.85&
245&
Mg\( \,  \)\textsc{xii}&
1.87&
\( 3.2\pm 0.8 \)\\
&
&
Mg\( \,  \)\textsc{xii}&
1.91&
\\
\hline 
\hline 
2.09&
10&
Si\( \,  \)\textsc{xiv}\textbf{}&
\textbf{2.01}&
-\\
&
&
Si\( \,  \)\textsc{xiii}&
2.17&
\\
\hline 
\hline 
&
&
Si\( \,  \)\textsc{xiii}&
2.29&
\\
2.37&
52&
Si\( \,  \)\textsc{xiii}&
2.38&
\( 5.9\pm 2.5 \)\\
&
&
S\( \,  \)\textsc{xv}\textbf{}&
\textbf{2.45}&
\\
\hline 
\hline 
&
&
Si\( \,  \)\textsc{xiv}&
2.50&
\\
2.56&
5&
Si\( \,  \)\textsc{xiv}&
2.60&
-\\
&
&
S\( \,  \)\textsc{xvi}\textbf{}&
\textbf{2.63}&
\\
\hline 
\hline 
&
&
S\( \,  \)\textsc{xv}&
2.88&
\\
2.99&
15&
S\( \,  \)\textsc{xv}&
3.02&
\( 4.0^{+5.5}_{-4.0} \)\\
&
&
Ar\( \,  \)\textsc{xvii}\textbf{}&
\textbf{3.08}&
\\
\hline 
\hline 
6.45&
1300&
Fe\( K\alpha  \)&
6.45&
\( 1.3\pm 0.1 \)\\
\hline 
\end{tabular}\par}\end{table}

This value for \( A_{Z} \) is still consistent with the Fe \( K\alpha  \)
line, which does not depend linearly with the abundance, but only logarithmically
\cite{mbf}, and it is strongly dependent on the other elements' abundance.
For \( A_{Z}\simeq 3 \), the EW of the iron line increases by only
a factor of \( \simeq 1.4 \), in agreement with the observations. Since there
are not detailed calculations of this effect for the other lines, we assume,
like \textsc{cloudy} apparently does, that their dependence on the abundance 
is almost linear; in any case, the derived \( A_{Z} \) should be considered
as a lower limit. It should be also noted that the model predict two other
weak line systems, centered at  at \( 2.09 \) and \( 2.56 \) \( keV \), which 
are not observed, not surprisingly given the very low equivalent widths. 

In the table we did not include the line at \( \simeq 7\, keV \), which could
be identified either as the \( K\beta  \) from neutral iron (\( 7.06\, keV \)),
or the line from Fe\( \,  \)\textsc{xxvi} (\( 6.97\, keV \)). Matt et al.
(1996) proposed a blend of these two lines, because they observed a very large
EW. This kind of solution would be unacceptable for our model, because it would
imply the presence of a highly ionized reflector for the Fe \( \,  \)\textsc{xxvi}
line to be produced. If we consider the \emph{SIS1} data, we face the same problem:
the EW of the line, if it has to be identified with the \( K\beta  \) only, should
be \( \simeq \frac{1}{9} \) the \( K\alpha  \), while the observed
one is much more intense, \( 0.3\pm 0.1 \). If, instead, we analyse the data
from \emph{SIS0}, the ratio \( K\alpha /K\beta  \) becomes consistent with
the expectations, being \( 0.15\pm 0.06 \). On the other hand,
\textsc{cloudy} does not seem to include the Fe \( K\beta  \) line in its calculations,
so we could not compare it with the data.

The comparison of the lines observed in the \emph{ASCA} spectrum 
at \( 0.82 \) and \( 1.34\, keV \) with the \textsc{cloudy} model is made
difficult by the large number of lines present, in the calculations, at
those energies. These lines are mainly $K$ lines from Ne and Mg and $L$
lines from Fe, but contributions from Ca, Mn and other elements are also
present. When folded with the $ASCA$ and $BeppoSAX$ response matrices, these
lines are blended to a pseudo--continuum, which contributes for about
10\% of the observed soft excess.
A detailed comparison between models and data for these lines
must await high resolution spectra, as those provided by the gratings 
onboard
{\it Chandra} and XMM--{\it Newton}.

Given the very simple assumptions, the agreement between observations and model
is remarkable. At the resolution of a CCD, therefore, a single,
mildly ionized material seems able to explain the Circinus X--ray spectrum. 

Finally, as a consistency check, we have also calculated the optical emission 
lines from such a reflector, and compared with those observed by Oliva, 
Marconi \& Moorwood
(1999) in the central, unresolved region of the source. 
As there may be other regions
contributing to the optical spectrum besides the X--ray reflector, 
we are contented
with not overproducing the observed lines. 
This goal is achieved assuming
a dust--to--gas ratio typical of the interstellar matter, and provided that
the temperature of the reflecting matter is larger than \( \sim  \)10\( ^{5} \)
K, otherwise the {[}O III{]}\( \lambda  \)5007 line is overproduced (an agreement
with this line is obtained for \( T\sim  \)2.5\( \times  \)10\( ^{5} \) K).
The calculated flux of the \( H\beta  \) is instead a few times lower than
observed, as for other intense lines (Tab. \ref{circ-opt}). It is worth noting
that dust is expected to be present, since the reflecting surface
of the torus is well beyond the evaporating radius, which 
is (Barvainis 1987):

\begin{equation}
\label{sublimation}
r_{ev}\simeq L^{1/2}_{46}\, pc\simeq 0.07\, pc
\end{equation}
 if we assume \( L_{tot}\simeq L_{IR}\simeq 5\times 10^{43}\, erg\, s^{-1} \)
(Siebenmorgen et al. 1997). It must be recalled
that dust may survive in an ionized medium 
provided that the electron temperature is lower than 10$^6$ K, see Draine \&
Salpeter (1979). 
The presence of dust, however, is in disagreement with the very low
$A_V$/$N_H$ ratio for this source. A possibility is that the size of the grains
is significantly larger than in the interstellar medium (Salvati \&
Maiolino 2000; Maiolino et al. 2000);
another possibility is that the optical/IR and X--ray absorbers are
physically distinct, which is well possible as the
optical and X--ray emission regions are certainly different.

\begin{table}

\caption{\label{circ-opt}Optical emission lines in Circinus 
(\protect\( H\beta =100\protect \)).}
{\centering \begin{tabular}{|c|c|c|c|}
\hline 
\textbf{\( \lambda  \) (Å)}&
\textbf{ESO NTT}&
\textbf{\textsc{cloudy}}\textbf{}&
\textbf{Id.}\\
\hline 
\hline 
6563&
298&
70&
\( H\alpha  \)\\
\hline 
5007&
1025&
260&
O\( \,  \)\textsc{iii}\\
\hline 
4959&
320&
90&
O\( \,  \)\textsc{iii}\\
\hline 
4861&
100&
26&
\( H\beta  \)\\
\hline 
4686&
41&
50&
He\( \,  \)\textsc{ii}\\
\hline 
3869&
136&
10&
Ne\( \,  \)\textsc{iii}\\
\hline 
\end{tabular}\par}\end{table}

\subsection{\label{1068_res}NGC~1068}

\subsubsection{The model}

Differently from the Circinus Galaxy, a single reflector
cannot explain the X--ray spectrum of NGC~1068. In fact, it is impossible
to have from a single region lines from neutral and He--like/H--like iron atoms.
Indeed, Guainazzi et al. (1999) showed that at least three reflectors are needed.
We used \textsc{cloudy} to derive the physical properties of these reflectors,
which in the following will be called the \textit{cold, warm} and \textit{hot}
reflectors.

The need for a third reflector arises from the presence of the O\( \,  \)\textsc{vii}
fluorescent line. It can be produced in the cold, thick reflector (which, as
in Circinus, can be naturally identified as the torus itself), but constraining
the ionization parameter to be too low to allow the production of all the other
lines, except for the `neutral' Fe \( K\alpha  \). 
On the other hand, in the hot reflector, responsible for the He-- and H--like
iron lines, all the lighter elements are 
fully stripped. The Mg, Si and S lines must therefore be produced in a third
reflector, the warm one.

\begin{figure*} 

\begin{minipage}{175mm}

\vspace{0.3cm}
{\par\centering \resizebox*{0.45\columnwidth}{!}{\includegraphics{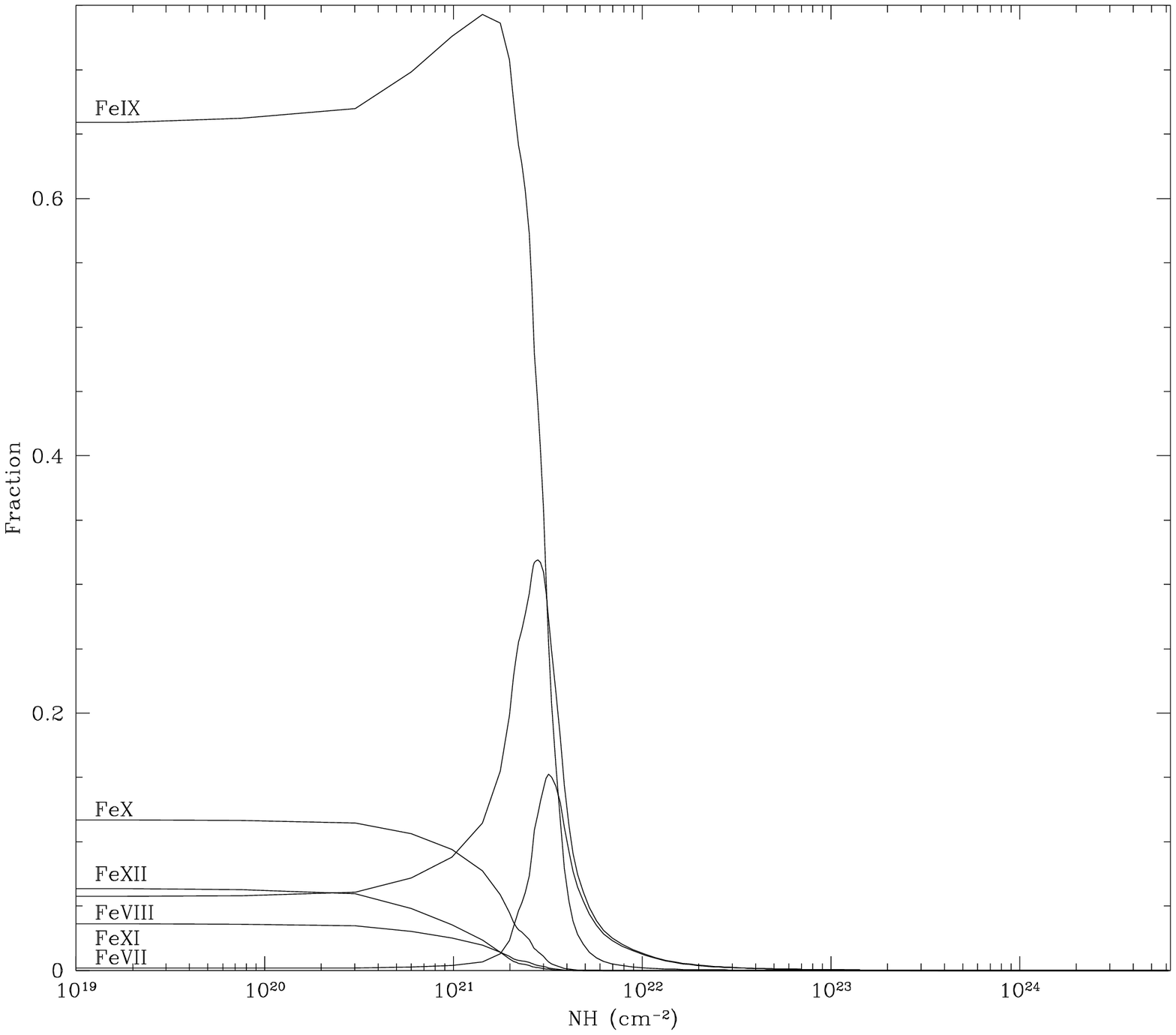}} 
\resizebox*{0.45\columnwidth}{!}{\includegraphics{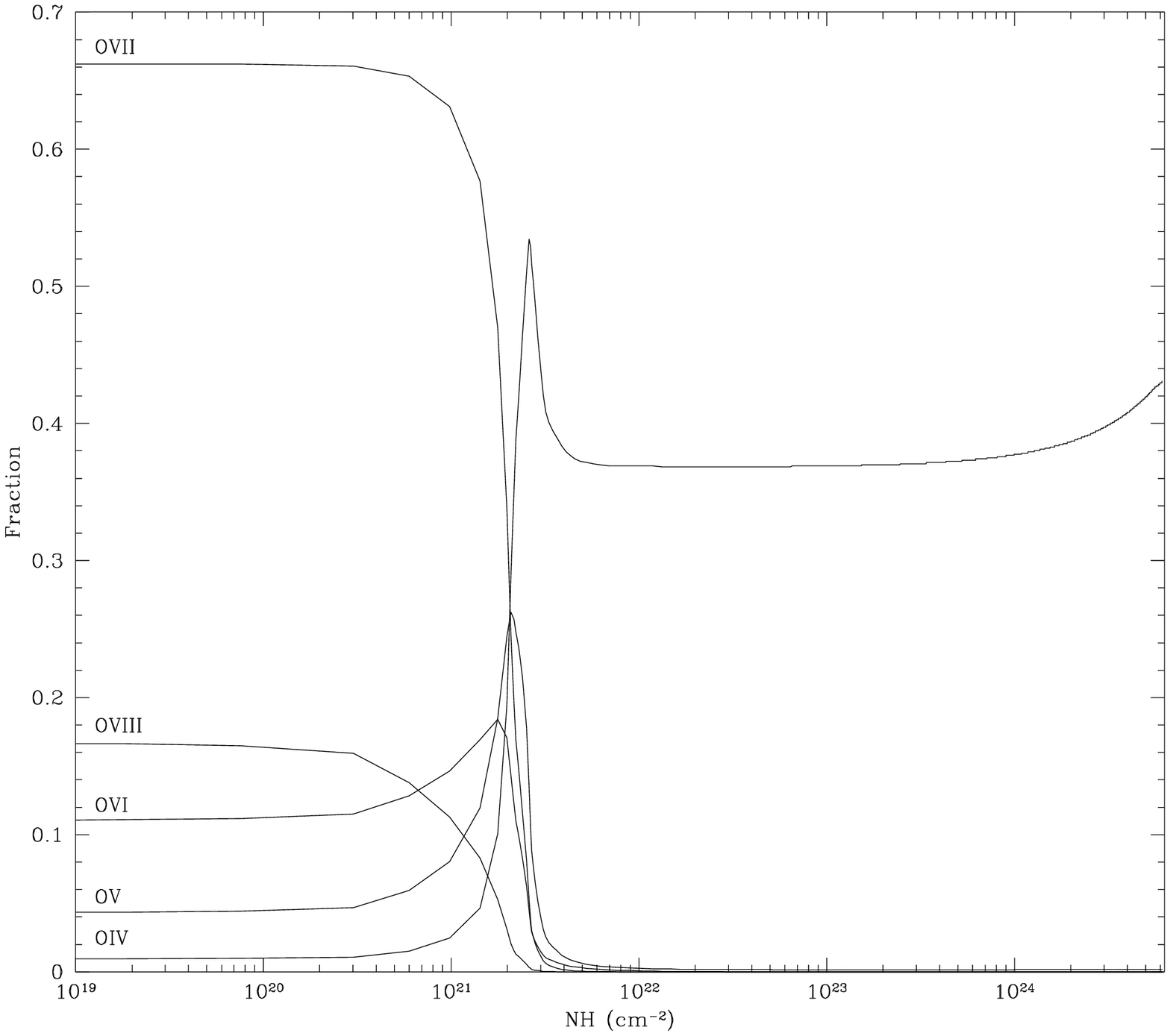}} \par}
\vspace{0.3cm}

\vspace{0.3cm}
{\par\centering \resizebox*{0.45\columnwidth}{!}{\includegraphics{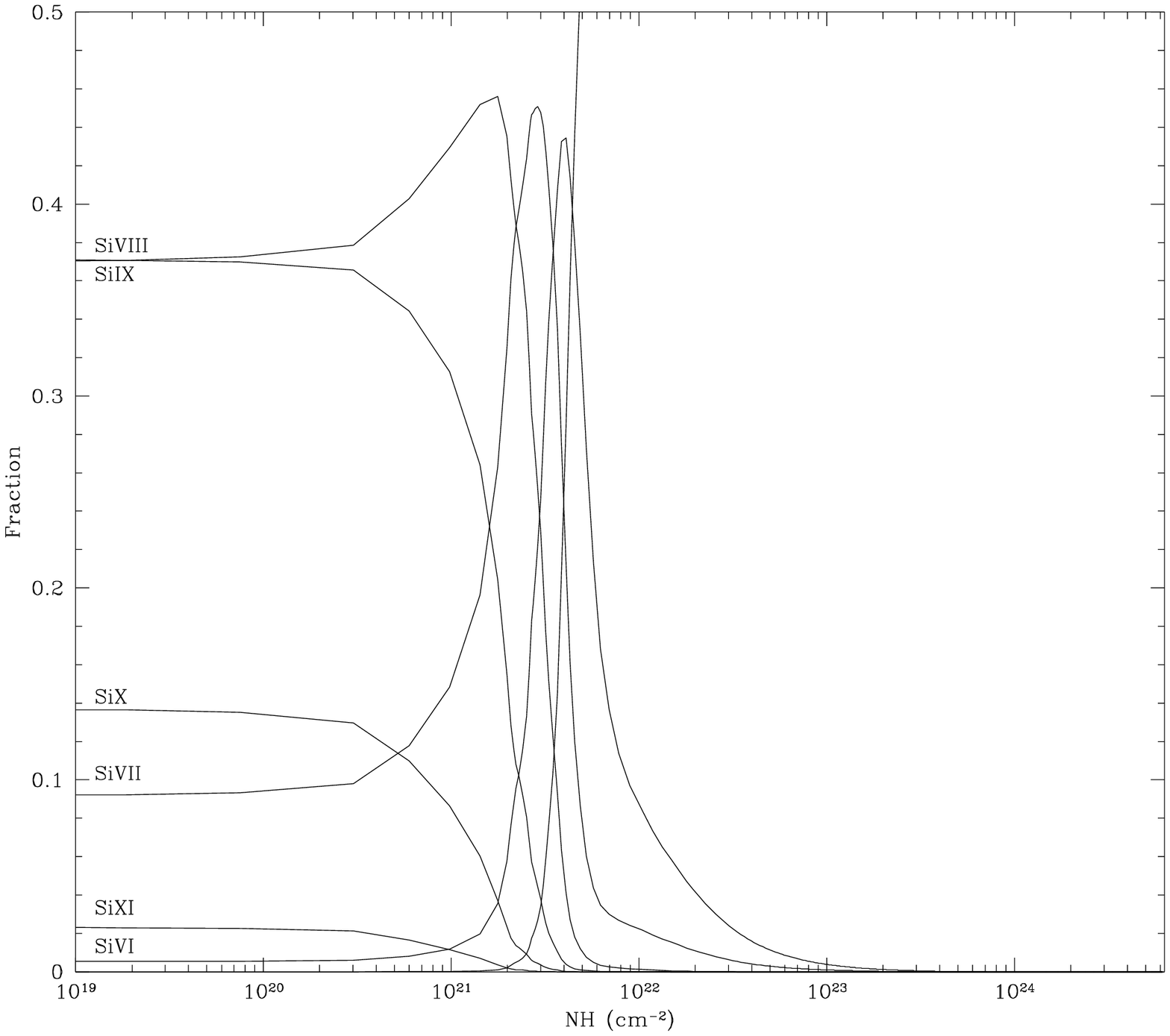}} 
\resizebox*{0.45\columnwidth}{!}{\includegraphics{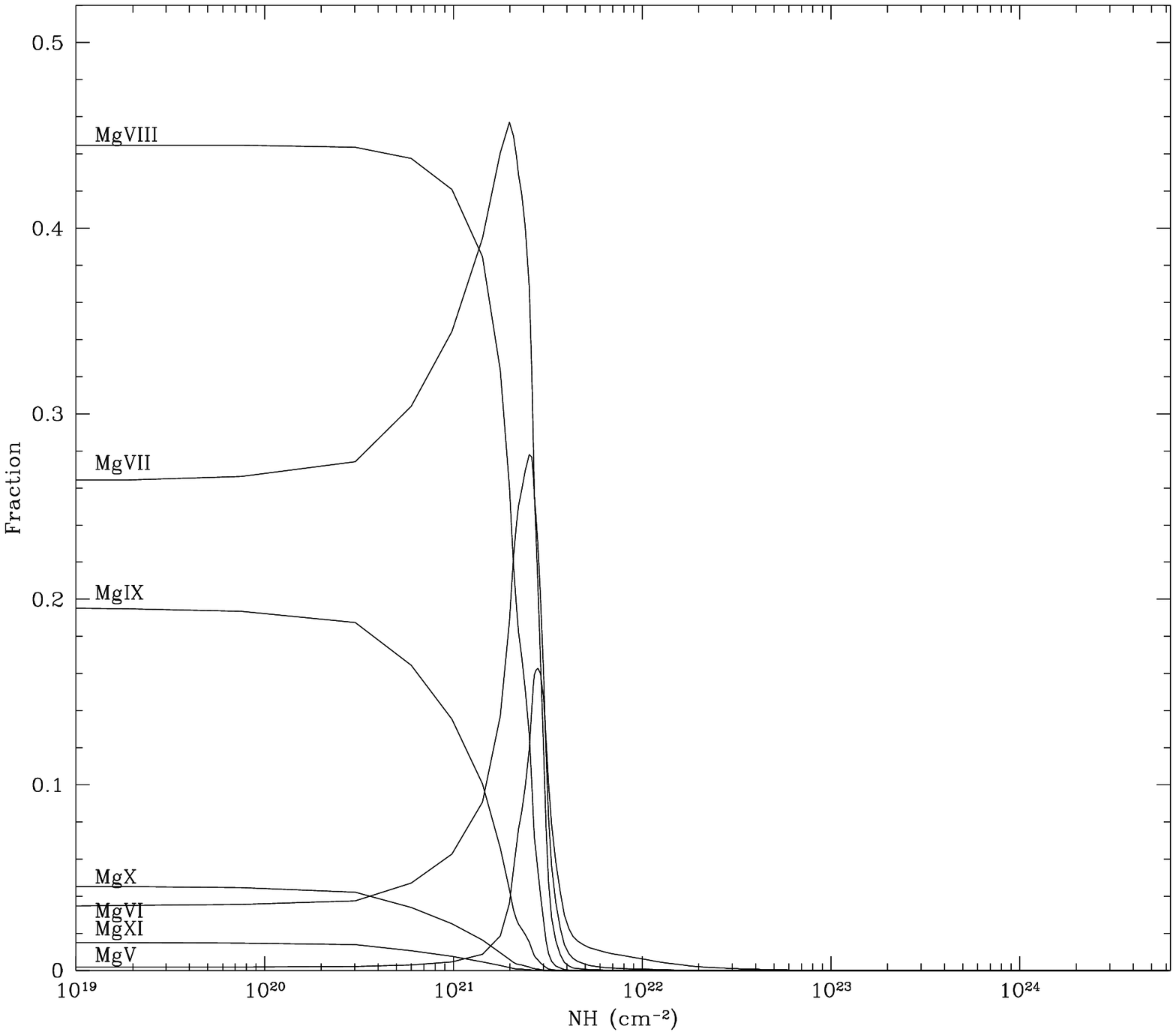}} \par}
\vspace{0.3cm}

\caption{Ionization structures in the cold reflector of NGC 1068 for Fe, O, 
Si and Mg. }

\label{1068coldioniz}

\end{minipage}

\end{figure*}

Let us investigate the characteristics of the three reflectors in more
detail. The illuminating radiation has been modeled as a power law 
with photon index 2.13 (Matt et al. 1997).
The cold reflector is made of low ionized material and can be identified with
the inner walls of the torus. We thus used a column density of \( 10^{26}\, cm^{-2} \)
(Matt et al. 1997) and an electron density of \( n_{e}=10^{6}\, cm^{-3} \),
corresponding to an outer radius of 30 pc.
The best ionization parameter is \( U=0.1 \), with a temperature of \( 10^{5}\, K \).
The ionization structures for some elements are shown 
in Fig. \ref{1068coldioniz}.

\begin{figure*} 

\begin{minipage}{175mm}

\vspace{0.3cm}
{\par\centering \resizebox*{0.45\columnwidth}{!}{\includegraphics{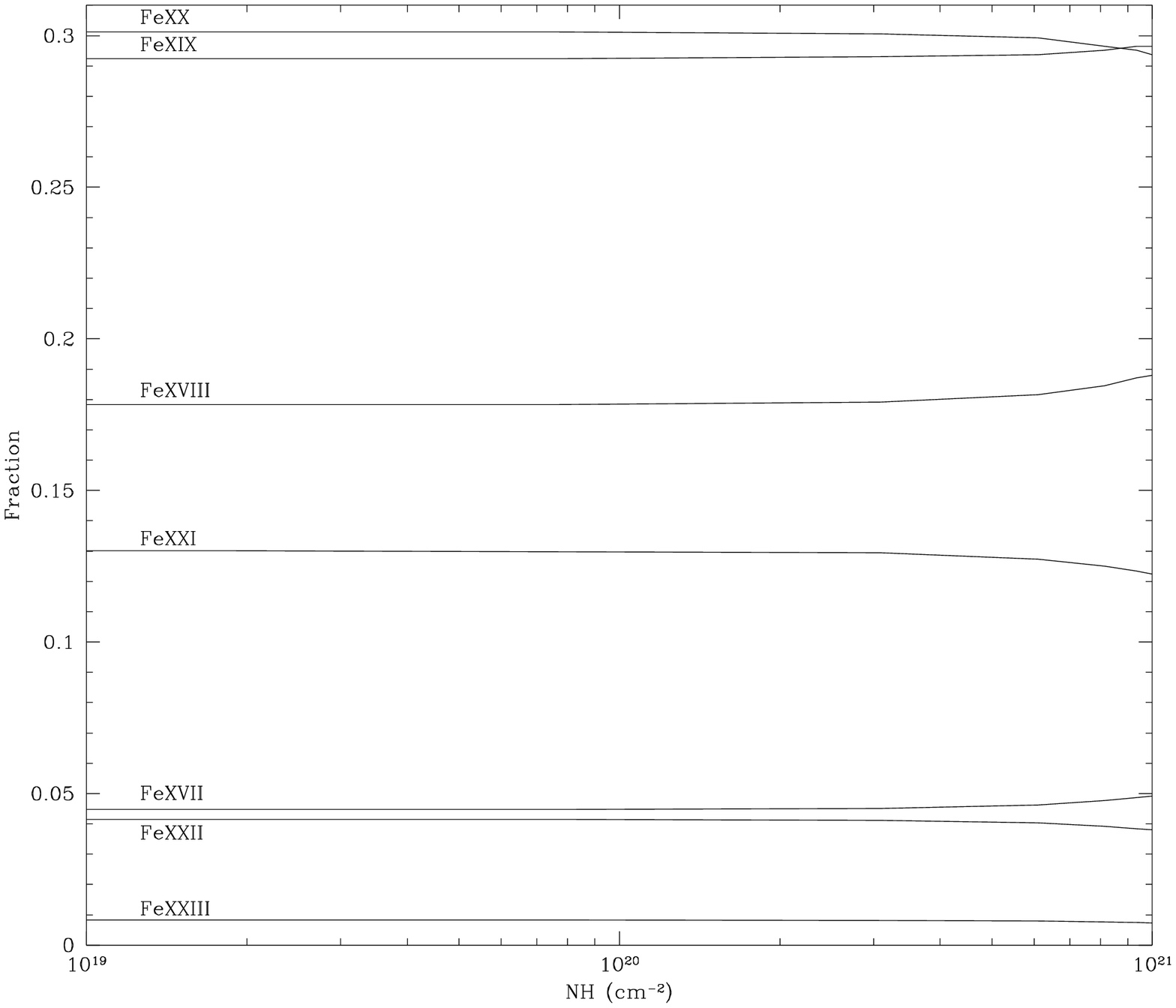}} 
\resizebox*{0.45\columnwidth}{!}{\includegraphics{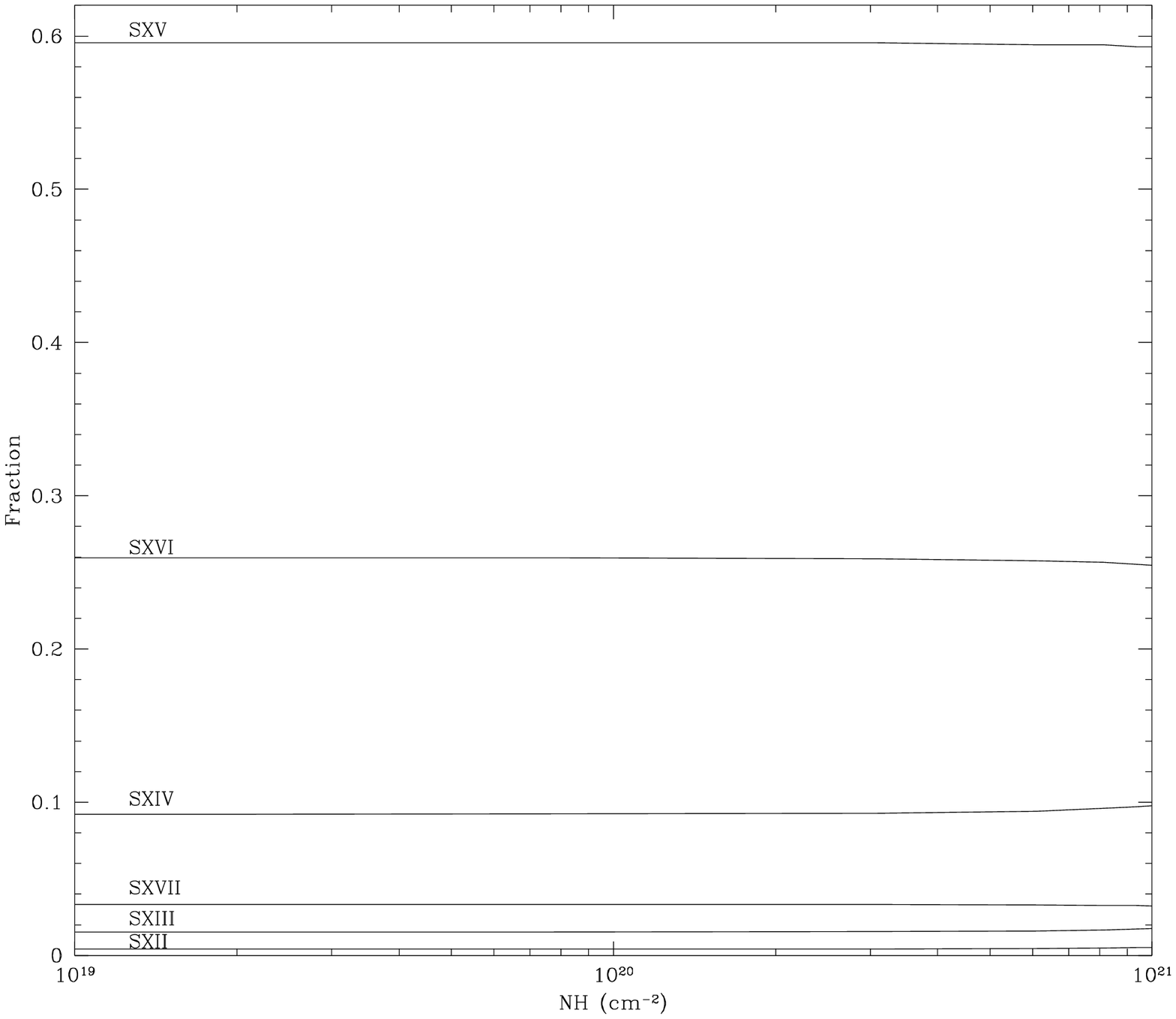}} \par}
\vspace{0.3cm}

\vspace{0.3cm}
{\par\centering \resizebox*{0.45\columnwidth}{!}{\includegraphics{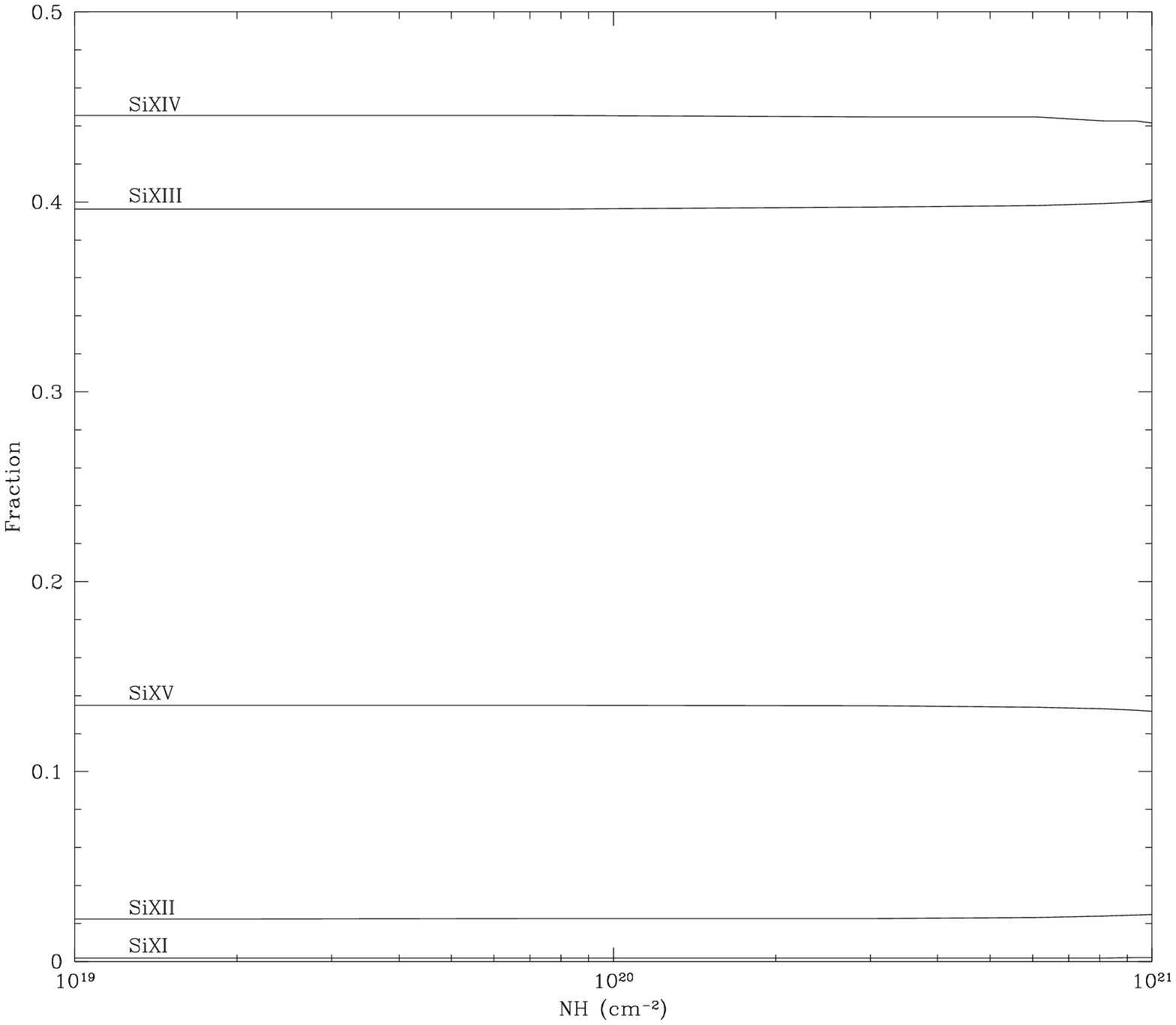}} 
\resizebox*{0.45\columnwidth}{!}{\includegraphics{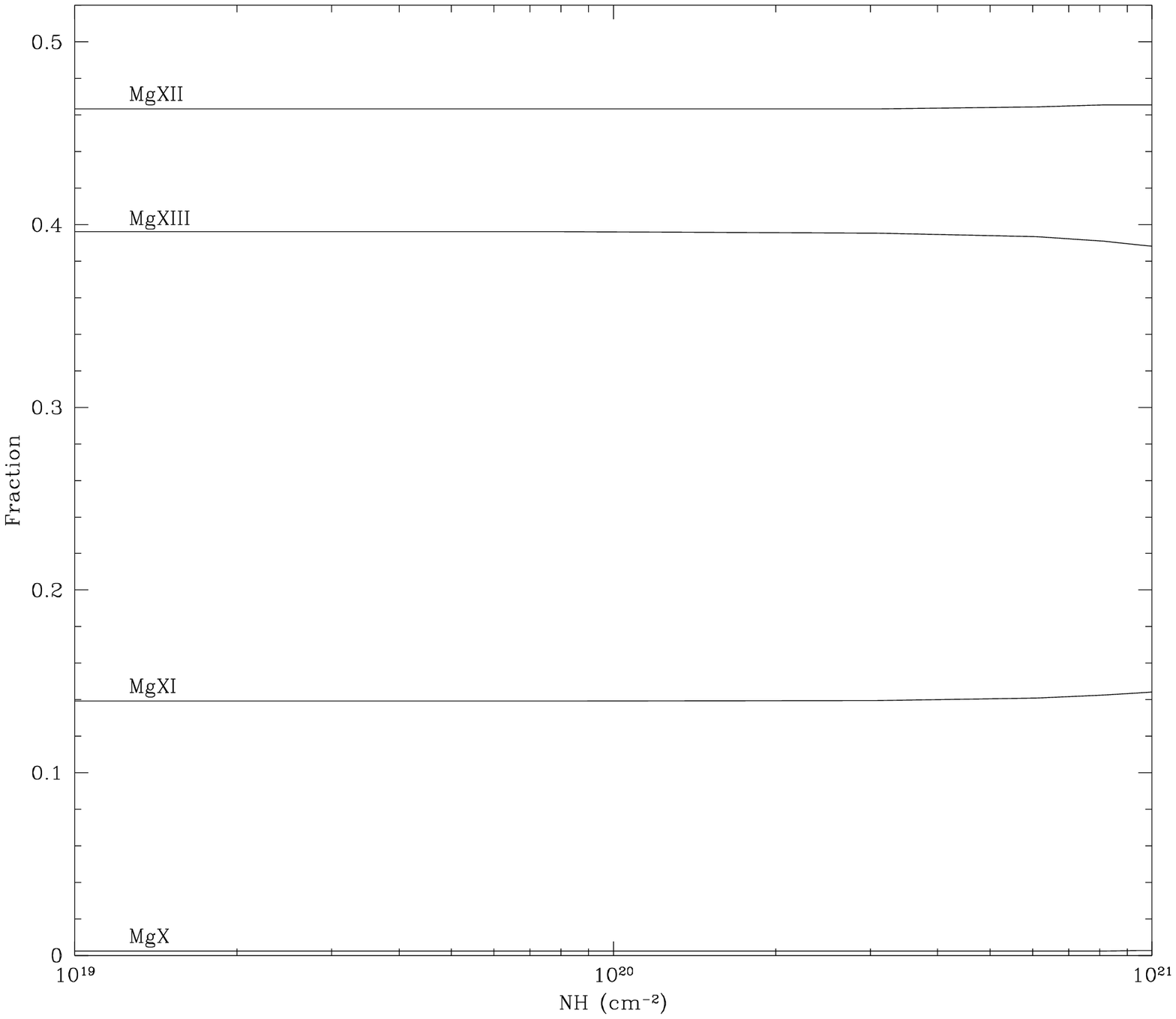}} \par}
\vspace{0.3cm}

\caption{Ionization structures in the warm reflector of NGC 1068 for Fe, S, 
Si and Mg. }

\label{1068warmioniz}

\end{minipage}

\end{figure*}

The warm reflector produces most of the remaining observed lines, except for
the Fe\( \,  \)\textsc{xxv} and Fe\( \,  \)\textsc{xxvi} lines. The best ionization
parameter is \( logU=0.7 \), and the temperature is \( 10^{6}\, K \). The ionization
structures for some elements are shown in Fig. \ref{1068warmioniz}: the oxygen
is fully stripped, while Fe\( \,  \)\textsc{xxv} and Fe\( \,  \)\textsc{xxvi}
have low abundances. We used a column density of \( 10^{21}\, cm^{-2} \) and
an electron density of \( n_{e}=10^{6}\, cm^{-3} \) (see below).

\begin{figure*} 

\begin{minipage}{175mm}

\vspace{0.3cm}
{\par\centering \resizebox*{0.45\columnwidth}{!}{\includegraphics{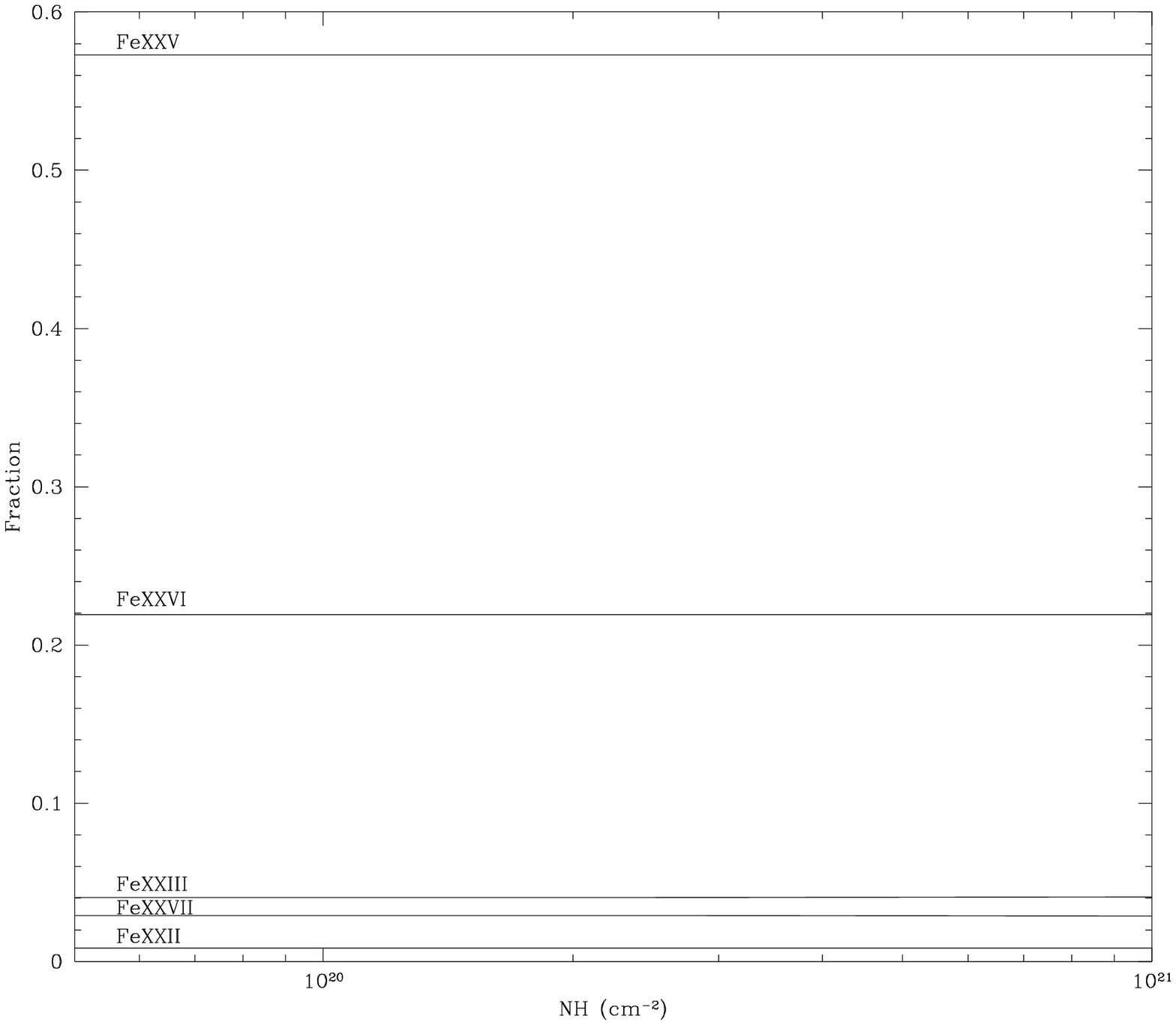}} 
\resizebox*{0.45\columnwidth}{!}{\includegraphics{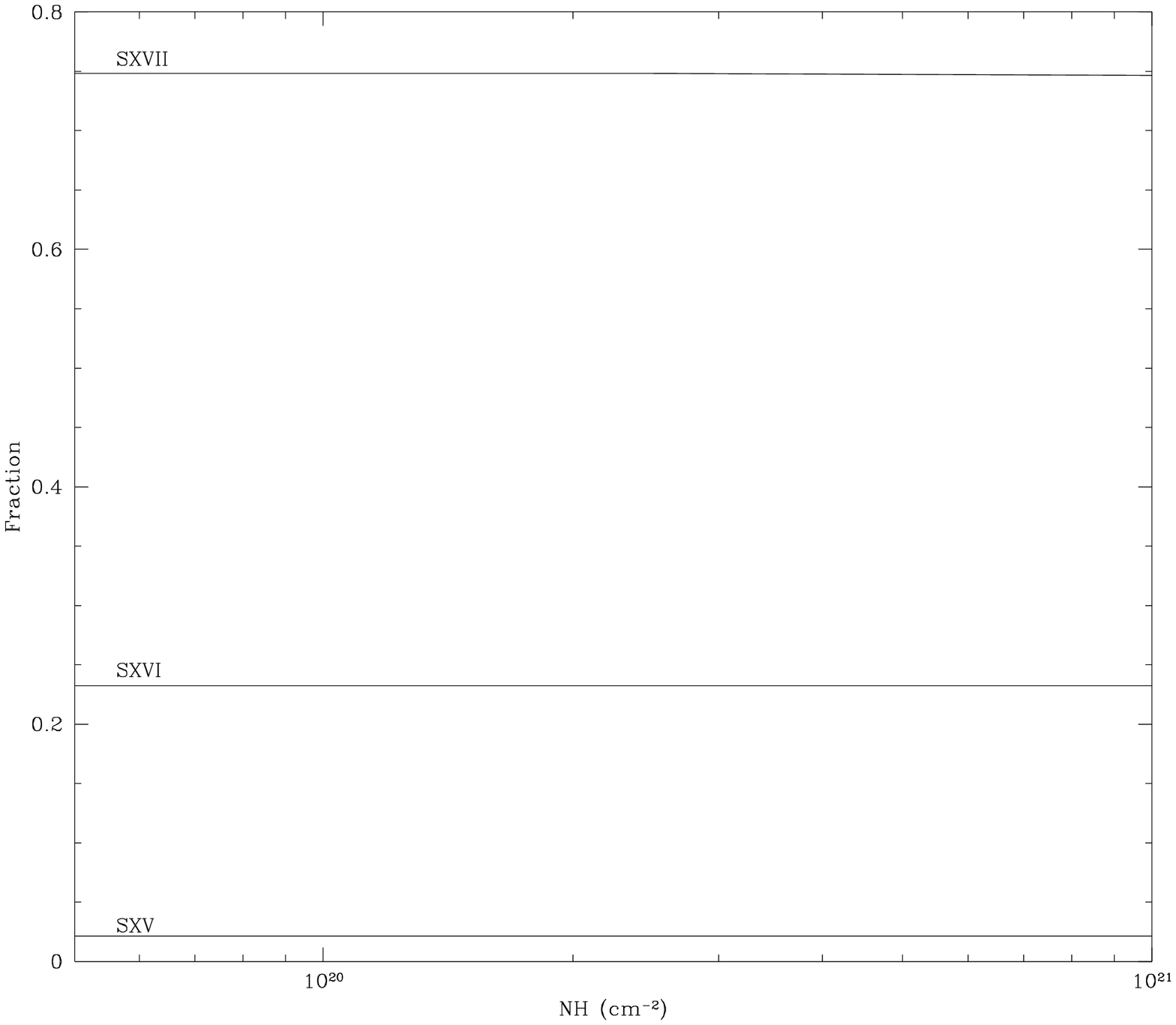}} \par}
\vspace{0.3cm}

\caption{Ionization structures in the hot reflector of NGC 1068 for Fe and S. }

\label{1068hotioniz}

\end{minipage}

\end{figure*}

The lines from the ionized iron come from the hot reflector, with a column 
density similar to that of 
the warm one but an ionization parameter of \( logU=2.1 \).
The ionization structure for some elements are shown in Fig. \ref{1068hotioniz}:
all elements but iron are fully stripped.

The values adopted for the column densities of the warm and the hot reflectors
have two justifications. The first constrain, which concerns the hot reflector
only, is from the EWs of the ionized iron
lines  which, as we will see in the next section, are in good agreement with
the results of Matt, Brandt \& Fabian (1996) with \( N_{H}\simeq 10^{21}\, cm^{-2} \).
The other constrain, which applies to the sum of the two reflectors, 
comes from the amount of the ionized reflection continuum,
considering its dependence from the nuclear luminosity of the source:

\begin{equation}
L_{r}\simeq L_{n}\cdot \tau _{C}\cdot f\leq L_{n}\cdot \tau _{C}
\end{equation}
 where \( f\equiv \frac{\Omega }{4\pi }\leq 1 \) is the covering factor and
\( \tau _{C}\equiv \frac{N_{H}}{1.5\cdot 10^{24}} \) is the Compton optical
depth. Assuming the observed \emph{ASCA} value for the warm reflection component,
\( L_{r}=0.6\cdot 10^{41}\, erg\, s^{-1} \), we can have an estimate of the
nuclear X--ray luminosity:

\begin{equation}
\label{luminosity}
L_{n}\geq \frac{9\cdot 10^{43}}{N^{21}_{H}}\, erg\, s^{-1}
\end{equation}
 Since the X-ray luminosity of NGC 1068 is estimated to be 
around \( 10^{44}\, erg\, s^{-1} \)
\cite{ifm} (corresponding to a bolometric luminosity of a few 
\( 10^{45}\, erg\, s^{-1} \) adopting the SED of Elvis et al. (1994), roughly
in agreement with the estimate of Pier et al. 1994), we see from 
(\ref{luminosity}) that we should use \( N_{H}\geq 10^{21}\, cm^{-2} \).
This is also the value adopted by \cite{g99}.

Finally, from the analysis of the EW of the lines it can be shown (see next
section) that the continua produced in the warm and in the hot reflectors 
should have column densities of the same order of magnitude, which justifies
the above assumptions.

We can estimate the inner radius of the torus in NGC 1068 similarly to what
have been done for Circinus. The difficulty here is that the nuclear luminosity
and the density of the matter are not known. With the best guess values
for these two quantities we have:

\begin{equation}
\label{NGCinnerradius}
r\simeq 3.9\, L^{1/2}_{{\rm X},44}n^{-1/2}_{6}\, pc
\end{equation}
For a discussion of the probable X--ray luminosity see Iwasawa et al. (1997).
For the density, we can only derive a lower limit. In fact:

\begin{equation}
\label{ngcdensity}
n_{e}=\frac{N_{H}}{R}>\frac{10^{25}}{R}\simeq 3\times 10^{4}\cdot \left( R_{10}\right) ^{-1}cm^{-3}
\end{equation}
 with the outer radius of the torus \( R \) in units of \( 10\, pc \). The (\ref{ngcdensity})
is a lower limit both because the column density could be grater
than \( 10^{25}\, cm^{-2} \), and because 
an upper limit on \( R \) of \( \simeq 100\, pc \)
(Risaliti et al. 1999) can be derived from the dynamical mass 
(Thatte et al. 1997) and
the mass of the gas itself measured from CO observations (Helfer 1997). 
 Our estimate for the inner radius is in agreement with the water
maser measurements of Gallimore et al. (1996), which have shown the presence 
of a keplerian disk with an inner radius \( r_{in}>1.3\, pc \) and
outer radius \( r_{out}\simeq 2.5\, pc \). 


Following the same considerations made for Circinus, we reanalysed \emph{ASCA}
and \emph{BeppoSAX} data using the \textsc{pexriv} model instead of \textsc{pexrav}.
The fits are of the same statistical quality as the previous ones, with no 
significant changes
in the values of the other parameters. The ionization parameter
which minimizes the \( \chi ^{2} \) in \emph{BeppoSAX} data is \( \xi =28_{-18}^{+16}\, erg\, cm\, s^{-1} \),
marginally consistent with the one calculated by 
\textsc{cloudy (}\( \xi \simeq 5\, erg\, cm\, s^{-1} \):
the conversion factor is different from the one we used in Circinus because
the power law slope is different).

Finally, we can estimate the visible fraction of the reflecting surface
in NGC 1068, comparing the fluxes of the Fe \( K\alpha  \) in our model with
the observed one. We obtain:

\begin{equation}
\phi \simeq 0.1
\end{equation}
using \( r\simeq 3.9\, pc \).

\subsubsection{\label{1068lines}The line spectrum}

Let us first discuss the iron line complex.
In Tab. \ref{NGCascaFelines} the Fe \( K \) lines from \emph{ASCA} data are
shown. All line energies were kept fixed in the fit, as well as the Fe \( K\beta  \)
flux (\( \frac{1}{9} \) of \( K\alpha  \)). The EWs are calculated against
the reflection continuum of the material that produces the line: Fe \( K\alpha  \)
and \( K\beta  \) against the cold reflection component (\textsc{pexra(i)v} model
in \textsc{Xspec}, cold reflector in \textsc{cloudy}), while Fe\( \,  \)\textsc{xxv}
and Fe\( \,  \)\textsc{xxvi} against the ionized component (power law in \textsc{Xspec},
hot reflector in \textsc{cloudy}).

\begin{table}

\caption{\label{NGCascaFelines}Fe K complex of NGC 1068 from \emph{ASCA} data. The
values with the symbol \protect\( ^{*}\protect \) were kept fixed during the
fit. The EWs are calculated against different continua, as explained in Sec.
\ref{1068lines}. }
{\centering \begin{tabular}{|c|c|c|}
\hline 
\textbf{E (keV)}&
\textbf{EW (eV)}&
\textbf{Ident}.\\
\hline 
\hline 
6.40\( ^{*} \)&
{\small \( 900_{-100}^{+130} \)} &
Fe \( K\alpha  \) `neutral'\\
\hline 
6.70\( ^{*} \)&
{\small \( 4350_{-750}^{+1000} \)} &
Fe\( \,  \)\textsc{xxv}\\
\hline 
6.96\( ^{*} \)&
{\small \( 940^{+940}_{-820} \)} &
Fe\( \,  \)\textsc{xxvi}\\
\hline 
7.06\( ^{*} \)&
\( 100^{*} \)&
Fe \( K\beta  \) `neutral'\\
\hline 
\end{tabular}\par}\end{table}

We compared the observational results for the iron lines with the work of
Matt, Brandt \& Fabian (1996) (from now on
MBF96), because it is specifically devoted to this problem,
while we continue using \textsc{cloudy} to calculate the relative abundances
of the iron ions. 

As discussed by MBF96, the Fe\( \,  \)\textsc{xxv} line is emitted 
both by 
resonant scattering and recombination; in the moderately thick case (for the resonant
line) of \( N_{H}=10^{21}\, cm^{-2} \), its EW can be expressed as:

\begin{equation}
\label{ewxxv}
EW_{FeXXV}\simeq (2100\cdot f_{XXV}\cdot A_{Z})+(1000\cdot f_{XXV}\cdot A_{Z})\, \, eV
\end{equation}
(see Fig. 5 of MBF96)
where \( f_{XXV}=0.57 \) is the ion abundance calculated by \textsc{cloudy.}
We must remember that this equivalent width has to be multiplied by a factor
\( X<1 \) (see previous section) to be compared with the observed one:

\begin{equation}
\label{feazx}
A_{Z}\cdot X=\frac{4350}{(3100\cdot 0.57)}=2.5
\end{equation}

The EW of the Fe\( \,  \)\textsc{xxvi} line can be expressed in a similar way:

\begin{equation}
EW_{FeXXVI}\simeq (1400\cdot f_{XXVI}\cdot A_{Z})+(400\cdot f_{XXVI}\cdot A_{Z})\, \, eV
\end{equation}
Using \( f_{XXVI}=0.22 \), as calculated by \textsc{cloudy}, and the estimate
of \( A_{Z}\cdot X \) (eq.\ref{feazx}), we have:

\begin{equation}
\label{ewxxvi}
EW_{FeXXVI}=990\, eV
\end{equation}
in good agreement with the observations. The value in (\ref{feazx})
is, therefore, consistent with both the ionized iron lines produced in the hot
reflector.

In Tab. \ref{NGCascalines} all the other lines observed by \emph{ASCA} are
shown. Almost all these lines are produced in the warm reflector: 
therefore, in the table are also reported the EWs multiplied by a factor \( 1-X=0.25 \)
which is our estimate for the contribution of the warm reflector alone (see
below). From (\ref{feazx}), this implies 
\( A_{Z}\simeq 3.3 \). This is the value used for
the metal abundances in our model: in this case, being the material optically thin,
the dependence of the EWs from \( A_{Z} \) is linear. Like
in Circinus, emission lines have
been blended when necessary: the results are summarized in Tab. \ref{NGCcloudy}.

\begin{table}

\caption{\label{NGCascalines}X-ray emission lines in NGC 1068 from \emph{ASCA} data
(\emph{SIS1} and \emph{SIS0}). The values obtained by dividing 
for by the factor \protect\( 1-X=0.25\protect \) (see text) are also shown. }
{\centering \begin{tabular}{|c|c|c|c|}
\hline 
\textbf{E (keV)}&
\multicolumn{2}{|c|}{ \textbf{EW (eV)}}&
\textbf{Ident}.\\
\cline{2-3} 
&
observed&
\( 1-X=0.25 \)&
\\
\hline 
\hline 
{\small \( 1.07_{-0.2}^{+0.2} \) }&
{\small \( 190_{-50}^{+190} \)}&
{\small \( 760^{+760}_{-200} \)} &
Fe\textsc{\( \,  \)l}\\
\hline 
{\small \( 1.20_{-0.2}^{+0.2} \)}&
{\small \( 210_{-60}^{+210} \)}&
{\small \( 840^{+840}_{-240} \)}&
Fe\textsc{\( \,  \)l}\\
\hline 
{\small \( 1.35_{-0.1}^{+0.1} \) }&
{\small \( 290_{-40}^{+30} \)}&
{\small \( 1160^{+120}_{-120} \)} &
Mg\textsc{\( \,  \)xi}\\
\hline 
{\small \( 1.81_{-0.1}^{+0.5} \) }&
{\small \( 330_{-30}^{+40} \)}&
{\small \( 1320^{+160}_{-120} \)} &
Si\textsc{\( \,  \)xiii}\\
\hline 
{\small \( 1.96_{-0.2}^{+0.3} \) }&
{\small \( 200_{-60}^{+50} \)}&
{\small \( 800^{+200}_{-240} \)}&
Si\textsc{\( \,  \)xiv}\\
\hline 
{\small \( 2.38_{-0.3}^{+0.2} \) }&
{\small \( 290_{-50}^{+60} \)}&
{\small \( 1160^{+240}_{-200} \)} &
S\textsc{\( \,  \)xv} - Si\textsc{\( \,  \)xiv}\\
\hline 
{\small \( 2.57_{-0.5}^{+0.4} \)}&
{\small \( 185_{-60}^{+80} \)}&
{\small \( 740^{+320}_{-240} \)}&
S\textsc{\( \,  \)xvi}\\
\hline 
{\small \( 3.10_{-0.5}^{+0.6} \) }&
{\small \( 150_{-60}^{+70} \)}&
{\small \( 600^{+280}_{-240} \)} &
Ar\textsc{\( \,  \)xvii}\\
\hline 
\end{tabular}\par}\end{table}

\begin{table}

\caption{\label{NGCcloudy}X-ray emission lines produced in our model for NGC 1068.
The energy in the first column is the mean value (weighted over the flux) of 
the lines that contribute to that blend. The lines with the symbol \protect\( ^{*}\protect \)
are contributes from the hot reflector. 
All the lines are \protect\( K\protect \)
transitions. The \protect\( K\alpha \protect \) lines which dominate each blend
are in boldface.}
{\centering \begin{tabular}{|c|c|cc|}
\hline 
\textbf{\( \overline{E} \)(keV)}&
\textbf{EW (eV)}&
\textbf{Id.}&
\textbf{E (keV)}\\
\hline 
\hline 
&
&
Si\textsc{\( \,  \)xiii}\textbf{}&
\textbf{1.85}\\
1.85&
1070&
Mg\textsc{\( \,  \)xii}&
1.87\\
&
&
Mg\textsc{\( \,  \)xii}&
1.91\\
\hline 
\hline 
2.01&
640 + 960\( ^{*} \)&
Si\textsc{\( \,  \)xiv}\textbf{}&
\textbf{2.01}\\
\hline 
\hline 
2.29&
860&
Si\textsc{\( \,  \)xiii}&
2.29\\
\hline 
\hline 
&
&
S\textsc{\( \,  \)xv}\textbf{}&
\textbf{2.45}\\
2.50&
1210&
Si\textsc{\( \,  \)xiv}&
2.50\\
&
&
Si\textsc{\( \,  \)xiv}&
2.55\\
\hline 
\hline 
2.60\( ^{*} \)&
1000&
Si\textsc{\( \,  \)xiv}&
2.60\\
\hline 
\hline 
3.11&
570&
Ar\textsc{\( \,  \)xvii}\textbf{}&
\textbf{3.08}\\
&
&
S\textsc{\( \,  \)xv}&
3.14\\
\hline 
\end{tabular}\par}\end{table}

Our choice for \( X \) is justified by the comparison of Tab. \ref{NGCcloudy}
with Tab. \ref{NGCascalines}. 
The choice
of \( X=0.75 \) and \( A_{Z}=3.3 \) is the one which gives a good agreeemnt
between model and observations. With these parameters, the comparison is 
satisfying apart from a deficit in the Si lines. Regarding 
the lines at energies lower than \( 1.85\, keV \), the 
model spectrum is too rich 
to allow for a useful comparison with the data, as already noted for Circinus.

The \textsc{O\( \,  \)vii} line is produced, as explained in the previous section,
by the cold reflector: its observed EW, calculated against the \textsc{pexriv}
continuum is \( 4000_{-1950}^{+2100}\, eV \). This value is in good agreement
with our model, which predicts \( 4000\, eV \), if we assume, as it seems reasonable,
\( A_{Z}=3.3 \) also for the cold reflector. Incidentally, it should be noted
again that the assumption of linear dependence of EWs from \( A_{Z} \) in optically
thick materials is not necessarily a good one.

The Fe \( K\alpha  \) at \( 6.4\, keV \) is also produced in the cold reflector.
The equivalent width measured by \emph{ASCA} (\( 900_{-100}^{+130}\, eV \))
is rather low when compared with the results of MBF96, taking also into accont
the metal overabundance discussed above, which implies a further increase
of the line flux by a factor of \( \simeq 1.4 \).
This correction contributes to make the EW of the `neutral' Fe \( K\alpha  \)
not consistent with \emph{ASCA} data: using a mean inclination angle, our model
predicts a value of \( \simeq 1600\cdot 1.4\simeq 2200\, eV \),  a
factor two larger than the observed one. On the other hand, \emph{BeppoSAX} data are
in agreement with \emph{ASCA}, giving an equivalent width just slightly higher
(\( \simeq 1100\, eV \)). This problem has no obvious solutions at present.

\section{Summary}

The results of our analysis may be summarized as follows:

For the Circinus Galaxy, a single, mildly ionized and optically thick
reflector can explain the line
spectrum and most of the continuum. Only the soft excess remains unexplained,
and it may be related to the starburst region. Recently, Sako et al. (2000)
have modeled the line spectrum of Circinus with the sum of a cold and thick 
plus an ionized and thin reflectors. At the spectral resolution of ASCA,
both models are tenable. We only note that our model is simpler, making use
of only  one, instead of two, reflectors. High resolution observations with
the gratings onboard {\it Chandra} and XMM--$Newton$ will certainly clarify
this issue.

In NGC~1068, instead, at least three reflectors are needed, one low, one mildly
and one highly ionized. (A thermal--like component, probably associated with 
the extended emission discovered by ROSAT/HRI, see Wilson et al. 1992, 
dominates in the soft X--ray band). In comparison with the Netzer \& Turner
(1997) model, our scenario for the line spectrum 
is necessarily more complex, because the detection of the 
O {\sc vii} line by $BeppoSAX$ requires one more reflector. On the other hand,
contrary to Netzer \& Turner(1997) and following MBF96 and Iwasawa, Fabian
\& Matt (1997), we identify the low ionized reflector with the very thick
absorber (Matt et al. 1997), so reducing by one the number of regions
needed to explain the overall spectrum. 

We identify the reflector in Circinus, and the low ionized one in NGC~1068,
with the inner
and visible part of the material responsible for the X--ray absorption. With
this assumption, we estimate for the inner radius of the absorber a value 
of 0.2 pc for Circinus and of a few parsecs for NGC~1068.

\section*{Acknowledgments}

We thank Roberto Maiolino, Alessandro Marconi and Fabrizio Nicastro for 
useful discussions, and the referee for useful comments and
suggestions. SB and GM acknowledges ASI and MURST (grant 
{\sc cofin}98--02--32) for financial support.

\end{document}